\documentclass[11pt]{iopart} 
\usepackage{epsfig,bm} 
\def \cD{{\cal D}} 
 
\def \cO{{\cal O}}

\def\bbbone{{\rm 1\hspace{-1.1mm}I}}	% unit matrix

\def \rd {\mbox{\rm d}}

\def \be{\begin{equation}}  
\def \ee{\end{equation}}  
\def \bea{\begin{eqnarray}}  
\def \eea{\end{eqnarray}}

\begin{document} 
%%%%%%%%%%%%%%%%%%%%%%%%%%%%%%%%%%%%%%%%%%%%%%%%%%%%%%%%%%%%%%%%%%%%%%%%%%%%%%%%%%%
\title{Spectra of Sparse Random Matrices} 
\author{Reimer K\"uhn}
\address{Mathematics Department, King's College London, Strand, London WC2R 2LS,UK\\[5mm] 
\today\\[-5mm] }
%%%%%%%%%%%%%%%%%%%%%%%%%%%%%%%%%%%%%%%%%%%%%%%%%%%%%%%%%%%%%%%%%%%%%%%%%%%%%%%%%%%
\begin{abstract}
We compute the spectral density for ensembles of of sparse symmetric random matrices 
using replica, managing to circumvent difficulties that have been encountered in 
earlier approaches along the lines first suggested in a seminal paper by Rodgers and 
Bray. Due attention is payed to the issue of localization. Our approach is not 
restricted to matrices defined on graphs with Poissonian degree distribution. 
Matrices defined on regular random graphs or on scale-free graphs, are easily handled.
We also look at matrices with row constraints such as discrete graph Laplacians.
Our approach naturally allows to unfold the total density of states into contributions 
coming from vertices of different local coordination. 
\end{abstract}
% PACS
% 02.50.-r 	Probability theory, stochastic processes, and statistics
% 05.10.-a 	Computational methods in statistical physics and nonlinear dynamics
%\maketitle 
%----------------------------------------------------------------------------------
\section{Introduction}
%----------------------------------------------------------------------------------
Since its inception by Wigner in the context of describing spectra of excited nuclei 
\cite{Wign51}, Random Matrix Theory (RMT) has found applications in numerous areas of
science, including questions concerning the stability of complex systems \cite{May72}, 
electron localisation \cite{Thoul77}, quantum chaos \cite{Bohigas+84},  Quantum
Chromo Dynamics \cite{Verbaar94}, finance \cite{Laloux+99,Plerou+99}, the physics 
of glasses  both at elevated \cite{Cav+99, Brod+01b} and low \cite{KuHo97,Kuehn03}
temperatures, number theory \cite{KeatingSnaith2000}, and many many more. For an 
extensive review  describing many of the applications in physics see, e.g. 
\cite{Guhr+98}.

In the present paper we revisit the problem of determining the spectral density for
ensembles of sparse random matrices pioneered two decades ago in seminal papers 
by Bray and Rodgers \cite{RodgBray88, BrayRodg88}. The problem has in recent years
received much renewed interest in connection with the study of complex networks,
motivated, for instance,  by the fact that geometric and topological properties of 
networks are reflected in spectral properties of adjacency matrices defining the 
networks in question \cite{AlbBarab02,Dorog+03}. Also, phenomena such as 
non-exponential relaxation in glassy systems  and gels \cite{BrayRodg88, Brod+01a}
--- intimately related to  Lifshitz tails \cite{Khor+06} and Griffiths' singularities 
in disordered systems  \cite{Griff69} --- as well as Anderson localization  of 
electronic \cite{EvangEcon92} or vibrational \cite{Cil+05} states have been 
studied in sparsely connected random systems, as finite dimensional versions of 
these problems have proven to be extremely difficult to analyse. A wealth of 
analytical and numerical results has been accumulated on these systems in recent 
years. Progress has, however, been partly hampered by the fact that full solutions  
of the Rodgers-Bray integral equation \cite{RodgBray88}, in terms of which spectral  
densities of the sparse random matrices in question are computed, have so far 
eluded us. Asymptotic analyses for large average connectivities \cite{RodgBray88,  
BrayRodg88}, and other approximation schemes such as the single defect approximation
(SDA) and the effective medium approximation (EMA) \cite{BirMon99,SemerjCugl02, 
Dorog+03} or very recently \cite{NagRog08}, as well as numerical diagonalization 
(e.g. \cite{Goh+01}) had to come in for help.

In what follows we describe some significant progress in the understanding of this
problem,  based upon advances in the statistical mechanical analysis of sparsely
connected  spin-glass like systems seen in the last couple of years \cite{Mon98,
MezPar01} --- in the present context in particular the proposal of a stochastic 
population-dynamics algorithm \cite{MezPar01} to solve the nonlinear integral 
equations appearing in the solution of these problems, and the recent generalization 
of these methods to systems with continuous degrees of freedom, such as models of
sparsely connected vector spins \cite{Coo+05}, or finitely coordinated models for
low-temperature phases of amorphous systems \cite{Ku+07}.

It is well known that the average spectral density of an ensemble ${\cal M}$ of 
$N\times N$ matrices $M$ can be computed from the ensemble average of the imaginary
part of their resolvent via
\be
\overline{\rho_N(\lambda)}= \frac{1}{\pi N}{\rm Im ~ Tr} ~
\overline{~\big[\lambda_{\varepsilon} \bbbone - M\big]^{-1}}\ ,
\label{avspec}
\ee
in which $\bbbone$ is the $N\times N$ unit matrix, and $\lambda_{\varepsilon} = 
\lambda - i\varepsilon$, the limit $\varepsilon  \to 0^+$ being understood. 
Following Edwards and Jones \cite{EdwJon76}, one can express this result in terms 
of the Gaussian integral
\be
Z_N = \int \prod_{i=1}^N \frac{\rd u_i}{\sqrt{2\pi/i}}~\exp\left\{
-\frac{i}{2}\sum_{i,j} u_i (\lambda_{\varepsilon} \delta_{ij} -M_{ij})u_j
\right\}
\label{Zn}
\ee
as
\be
\overline{\rho_N(\lambda)} = -\frac{2}{\pi N}~{\rm Im}~
\frac{\partial}{\partial \lambda} ~\overline{\ln Z_N} = 
\frac{1}{N}{\rm Re} ~ \sum_{i=1}^N \overline{\langle u_i^2 \rangle}\ ,
\label{avlog}
\ee
using the replica method to evaluate the average of the logarithm in (\ref{avlog})
over the ensemble ${\cal M}$  of matrices $M$ under consideration.  The 
`averages' $\langle u_i^2 \rangle$ in (\ref{avlog}) are evaluated with respect to
the `Gaussian measure' defined by (\ref{Zn}).\footnote{Note that we are using 
probabilistic notions in a loose, metaphorical sense, as the Gaussian measures 
used in these calculations are complex.} This has been the path taken in 
\cite{RodgBray88}; we shall initially follow their reasoning.

Disregarding the complex nature of the `Hamiltonian' in the evaluation of 
(\ref{Zn}), the mathematical problem posed in (\ref{Zn}), (\ref{avlog}) is analogous 
to the evaluation of an `internal energy´ of a disordered system with quenched disorder. 
Within the general class of finitely coordinated amorphous model systems considered 
in \cite{Ku+07}, the one represented by (\ref{Zn}), (\ref{avlog}) constitutes a 
particular sub-class, viz. that of harmonically coupled systems, for which the 
analysis was found to be {\em much\/} simpler than for systems involving anharmonic 
couplings. Indeed, while the solution of the latter required the self-consistent 
determination of probability distributions over infinite dimensional function-spaces, 
it was realized in \cite{Ku+07} that solutions of harmonically coupled systems 
could be formulated in terms of superpositions of Gaussians, and that the 
self-consistency problem reduced to the (much simpler) problem of a self-consistent 
determination of the probability distribution of their variances. 

It can be fairly argued that this last insight is, in fact, easier to obtain within
a Bethe-Peierls or cavity type approach \cite{MezPar01}, in which (\ref{Zn}) is 
recursively evaluated {\em for given instances\/} on graphs which are locally tree-like, 
ignoring correlations among subtrees --- an approximation that becomes exact, e.g., 
for random graphs that remain finitely coordinated in the thermodynamic limit. This 
approach is taken in a separate publication \cite{Rog+08}, in which (finite)
single-instances and promising algorithmic aspects of the problem are being 
highlighted. 

Although \cite{Ku+07} describes all technical details needed for a replica analysis
of the present problem, we shall nevertheless reproduce the key steps here, both to 
keep the paper self-contained, and to point out along the way where the impasse in
\cite{RodgBray88} arises, and how it is circumvented.

The remainder of the paper is organized as follows. In Sec. 2, we describe the replica
analysis of the problem posed by (\ref{Zn}), (\ref{avlog}), specializing to matrices
defined on Poissonian (Erd\"os-Renyi) random graphs. It has been known for some time
\cite{EdwJon76,RodgBray88} that the replica-symmetric high-temperature solution ---
i.e., a solution preserving both, permutation-symmetry among replica, and rotational 
symmetry in the space of replica --- is exact for problems of the type considered here.
Accordingly, a representation that respects these symmetries is formulated in Sec. 2.1.
It is at this point where our formulation departs from that of \cite{RodgBray88}. In
Sec. 3 we present results for a variety of examples, and compare with numerical 
diagonalization results for large finite matrices to assess their quality. In 
sufficiently sparse graphs, one expects localized states to appear.  The signatures 
of localization within our approach are discussed throughout Sec. 3, with inverse 
participation ratios (IPRs) as a diagnostic tool looked at in Sec. 3.2. A detailed 
investigation of Anderson localization for (discrete) Schr\"odinger operators on 
sparse random graphs will be reserved to a separate publication \cite{KuMour08}. 
Matrices with bimodal instead of Gaussian random couplings are studied in Sec. 3.3.
As the formal structure of the self-consistency problem remains unaltered when the 
Poissonian random graphs are replaced by graphs with other degree distributions 
\cite{Ku+07}, we can exploit this fact to present results for regular and scale-free 
random graphs in Sec. 3.4. Modifications needed to treat matrices with row-constraints, 
such as discrete graph Laplacians are outlined in Sec. 3.5. Our approach naturally 
allows to unfold the total density of states into contributions coming from vertices 
of different local coordination, and we finally present an example of such an 
unfolding in Sec. 3.6. The final Sec. 4 contains a brief summary and an outlook on 
promising directions for future research.

%----------------------------------------------------------------------------------
\section{Replica Analysis}
%----------------------------------------------------------------------------------
%----------------------------------------------------------------------------------
\subsection{General Formulation}
%----------------------------------------------------------------------------------
Here we briefly outline the evaluation of (\ref{Zn}), (\ref{avlog}) for sparse 
symmetric matries $M$ of the form
\be
M_{ij} = c_{ij} K_{ij} \ ,
\ee
in which $C= \{c_{ij}\}$ is a symmetric adjacency matrix of an undirected random 
graph (with $c_{ii}=0$), and the non-zero elements of $M$ are specified by the  
$K_{ij}$, also taken to be symmetric in the indices. Within the present outline 
we restrict ourselves for the sake of simplicity to adjacency matrices of 
Erd\"os-Renyi random graphs, with
$$
P(\{c_{ij}\}) = \prod_{i<j} p(c_{ij}) \delta_{c_{ij},c_{ji}}\qquad {\rm and}\qquad  
p(c_{ij})= \left(1- \frac{c}{N}\right) \delta_{c_{ij},0}+ \frac{c}{N} \delta_{c_{ij},1}\ ,
$$
exhibiting a Posisssonian degree distribution with average coordination $c$.
We note at the outset that formal results carry over without modification to other 
cases  \cite{Ku+07}. There is no need at this point to specify the distribution of the 
$K_{ij}$, but we shall typically look at Gaussian and bimodal distributions. 

The average (\ref{avlog}) is evaluated using replica $\overline{\ln Z_N} = 
\lim_{n\to 0}\, \frac{1}{n}\, \ln \overline{Z_N^n}$, starting with integer numbers
of replica as usual. After performing the average over the distribution of the
connectivities one obtains
\be
\hspace{-20mm} \overline{ Z_N^n} = \int \prod_{ia} \frac{\rd u_{ia}}{\sqrt{2\pi/i}} 
\exp\left\{ -\frac{i}{2}\lambda_\varepsilon \sum_{i,a} u_{ia}^2 
+ \frac{c}{2N} \sum_{ij} \left(\left\langle\exp\Bigg(i K \sum_{a} u_{ia}u_{ja}\Bigg)
\right\rangle_K 
 -1 \right)\right\}\ ,
\label{avZn}
\ee
in which $\langle \dots\rangle_K$ refers to an average over the distribution of 
the $K_{ij}$. A decoupling of sites is achieved by introducing the replicated 
density
\be
\rho(\bm u) =\frac{1}{N} \sum_i \prod_a \delta\Big( u_a - u_{ia}\Big)\ ,
\ee
with $\bm u$ denoting the replica vector $\bm u = (u_1, u_2, \dots , u_n)$,
and enforcing its definition via functional $\delta$ distributions, 
\be
1 = \int \cD \rho \cD \hat \rho~\exp\left\{-i \int \rd \bm u \hat \rho(\bm u) 
\Bigg(N \rho(\bm u) - \sum_i\prod_a \delta\Big( u_a - u_{ia}\Big)\Bigg)
\right\}\ .
\ee
This gives (using shorthands of the form  $\rd \rho(\bm u) \equiv \rd \bm u \rho(\bm u)$ 
where useful)
\bea
\hspace{-10mm} \overline{Z_N^n} &=& \int \cD \rho \int \cD \hat\rho ~
\exp\left\{N\left[
  \frac{c}{2}\int \rd \rho(\bm u) \rd \rho(\bm v) 
  \left(
   \left \langle\exp\Bigg(i K \sum_{a} u_{a} v_{a}\Bigg)\right\rangle_K  -1
  \right) 
\right.\right. \nonumber\\
&& \left.\left. 
  -\int \rd \bm u \, i \hat\rho(\bm u)\rho(\bm u) 
 + \ln \int \prod_a \frac{\rd u_a}{\sqrt{2\pi/i}} \exp\Bigg(i\,\hat\rho(\bm x) -
\frac{i}{2}\,\lambda_\varepsilon \sum_a u_a^2 \Bigg)
\right] \right\}\ ,
\label{avZn2}
\eea
allowing to evaluate $N^{-1}\ln \overline{Z_N^n}$ by a saddle point method. 
The stationarity conditions w.r.t. variations of $\rho$ and $\hat\rho$ read
\bea
 i\hat\rho(\bm u) &=& c\int \rd \rho(\bm v) 
\left(\left\langle\exp\Bigg(i K \sum_{a} u_{a} v_{a}\Bigg)\right\rangle_K  -1\right)\ ,
\label{fpehatrho}\\
 \rho(\bm u) &=&
\frac{\exp\Bigg(i\,\hat\rho(\bm u) - \frac{i}{2}\,\lambda_\varepsilon \sum_a u_a^2 \Bigg)}
{ \int  \rd \bm u \exp\Bigg(i\,\hat\rho(\bm u) -
\frac{i}{2}\,\lambda_\varepsilon \sum_a u_a^2 \Bigg)}\ .
\label{fperho}
\eea
The way in which sites are decoupled constitutes the first point of departure between
our treatment and that of \cite{RodgBray88} and subsequent analyses inspired by it (e.g.
\cite{Rodg+05,KimKahng07}). In these papers the averaged exponential expressions in
the exponent of (\ref{avZn}),
\be
f(\bm u_i\cdot \bm v_j)= f\Big(\sum_{a} u_{ia} v_{ja}\Big) = \left\langle\exp\Big(i K 
\sum_{a} u_{ia} v_{ja}\Big)\right\rangle_K  -1\ ,
\ee
is expanded, and an infinite family of multi-replica generalizations of Edwards Anderson 
order parameters (and corresponding Hubbard-Stratonovich transformations) are used to 
decouple the sites, much as in the treatment of the dilute spin-glass problem by Viana 
and Bray \cite{ViaBray85}. The authors then use the expansion and the infinite set of self-consistency equations for the multi-replica generalizations of Edwards Anderson 
order parameters to construct a non-linear integral equation for a function $g$ defined 
via a suitable `average' of $f$; see \cite{RodgBray88} for details. Our treatment in 
this respect is closer in spirit to the alternative approach of Kanter and Sompolinsky 
\cite{KantSomp87b} who treat local field distributions (which in the general context of 
disordered amorphous systems discussed in \cite{Ku+07} become distributions of local 
potentials) as the primary object of their theory. 

However, the difference between our treatment and that of \cite{RodgBray88} is at this 
point still superficial. Indeed, we have the correspondence 
\be
 i\hat\rho(\bm u) = c g(\bm u)
\ee
between our `conjugate density' $\hat\rho$ and the function $g$ of \cite{RodgBray88}.
With this identification, (\ref{fpehatrho}) and (\ref{fperho}) can be combined to give
\be
g(\bm u) = \frac{\int \rd \bm v ~f(\bm u\cdot \bm v)~\exp\Big(c g(\bm  v) - \frac{i}{2}\,\lambda_\varepsilon \bm v^2\Big) }{\int \rd \bm v ~\exp\Big(c g(\bm  v) - \frac{i}{2}\,\lambda_\varepsilon \bm v^2\Big)}\ ,
\label{fpeg}
\ee
which is the Rodgers-Bray integral equation for general distributions of non-zero 
bond strengths.
%----------------------------------------------------------------------------------
\subsection{Replica Symmetry}
%----------------------------------------------------------------------------------
To deal with the $n\to 0$ limit in these equations, assumptions concerning the 
invariance properties of the solutions $\rho(\bm u)$ and $\hat\rho(\bm u)$  of 
(\ref{fpehatrho}) and (\ref{fperho}) --- alternatively of the solution $g(\bm u)$ 
of (\ref{fpeg})--- under transformations among the replica are required. It has been 
established for some time \cite{EdwJon76,RodgBray88} that the replica-symmetric 
high-temperature solution --- i.e., a solution preserving both, permutation-symmetry 
among replica, and rotational symmetry in the space of replica --- is exact for 
problems of the type considered here. It is here where the paths taken in the 
present paper and in \cite{RodgBray88} really bifurcate. In \cite{RodgBray88},
the assumption $g(\bm u)=g(u)$, with $u=|\bm u|$ is used to perform the angular 
integrals in $n$-dimensional polar coordinates in (\ref{fpeg}), resulting in an 
integral equation for $g(u)$ in the $n\to 0$-limit. This integral equation has
also been obtained using the supersymmetry approach \cite{FyodMirl91}. It has, 
however, so far resisted exhaustive analysis or full numerical solution. 

In the present paper we follow \cite{Ku+07}, and represent $\rho$ and $\hat\rho$
as superpositions of replica-symmetric functions, using the observation made in
\cite{Ku+07} that superpositions of Gaussians of the form
\bea
\rho(\bm u) &=& \int \rd\pi(\omega) \prod_a 
\frac{\exp\big[-\frac{\omega}{2} u_a^2\big]}{Z(\omega)}\ ,
\label{reprho}\\
i\hat\rho(\bm u) &=& \hat c \int \rd \hat\pi(\hat{\omega}) \prod_a 
\frac{\exp\big[-\frac{\hat\omega}{2}u_a^2 \big]}{Z(\hat{\omega})}\ ,
\label{rephatrho}
\eea
would provide exact solutions for harmonically coupled systems. Note that these
expressions do indeed preserve permutation symmetry among replica as well 
as rotational symmetry. In (\ref{rephatrho}) the constant $\hat c$ is to be 
determined such that $\hat\pi$ is normalized, $\int \rd \hat\pi(\hat{\omega}) 
= 1$. We note that these representations make sense only for ${\rm Re}~\omega 
>0$ and ${\rm Re}~\hat \omega >0$; later on we shall find that these conditions 
are self-consistently met for solutions of the fixed point equations. Expressing 
(\ref{avZn2}) in terms of $\pi$ and $\hat\pi$, we get
\be
\overline{Z_N^n } = \int \cD \pi\cD \hat \pi  \exp\left\{N \left[G_1[\pi] + 
G_2[\hat\pi,\pi] + G_3[\hat\pi]\right]\right\}\ .
\label{ZnGi}
\ee
As $n\to 0$, the functionals $G_1$, $G_2$ and $G_3$ evaluate to
\bea
G_1[\pi] &\simeq & n \frac{c}{2} \int\rd \pi(\omega)\rd\pi(\omega')~
\left\langle\ln \frac{Z_2(\omega,\omega',K)}{Z(\omega)Z(\omega')}\right\rangle_K\ ,\\
G_2[\hat\pi,\pi] &\simeq & - \hat c - n \hat c\int \rd \hat\pi(\hat{\omega}) 
\rd\pi(\omega)~ \ln \frac{Z(\hat{\omega}+\omega)}
{Z(\hat\omega)Z(\omega)}\ ,\\
G_3[\hat\pi] &\simeq & \hat c   + n~
\sum_{k=0}^\infty p_{\hat c}(k) \int \{\rd \hat\pi\}_k ~
\ln \frac{Z_\lambda(\{\hat\omega\}_k)}{\prod_{\ell=1}^k Z(\hat\omega_\ell)}\ ,
\eea
in which we have introduced the shorthands $\{\rd \hat\pi\}_k \equiv \prod_{\ell=1}^k 
\rd \hat\pi(\hat\omega_\ell)$, and $\{\hat\omega\}_k= \sum_{\ell=1}^k \hat\omega_\ell$, 
a Poissonian connectivity distribution
\be
p_{\hat c}(k) = \frac{\hat c^k}{k!} \exp[-\hat c]
\ee
with average connectivity $\langle k\rangle = \hat c$, and the `partition functions'
\bea
\hspace{-5mm} Z(\omega) &=& \int \rd u ~\exp\left[-\frac{\omega}{2} u^2\right]
= \sqrt{2 \pi/\omega}\ ,
\label{defZ1s}\\
\hspace{-5mm} Z_{\lambda_\varepsilon}(\{\hat\omega\}_k)  &=& \int \frac{\rd u}{\sqrt{2\pi/i}}~
\exp\left[-\frac{1}{2}\bigg(i\lambda_\varepsilon+ \{\hat\omega\}_k\bigg) u^2 \right] 
=\left(\frac{i}{i\lambda_\varepsilon+ \{\hat\omega\}_k}\right)^{1/2}\ ,
\label{defZs}\\
\hspace{-5mm} Z_2(\omega,\omega',K) &=& \int \rd u \rd v ~
\exp\left[-\frac{1}{2}\bigg(\omega u^2 + \omega' v^2 - 2i K u v\bigg)\right] 
= \frac{2\pi}{\sqrt{\omega \omega' + K^2}}\ .
\label{defZ2s}
\eea
Note that the $\cO(1)$ contributions of $G_2$ and $G_3$ in the exponent
of  (\ref{avZn2}) cancel in their sum.

The stationarity condition  of the functional integral (\ref{avZn2}) w.r.t
variations of $\rho$ and $\hat\rho$ is reformulated in terms of stationarity 
conditions w.r.t variations $\pi$ and $\hat\pi$,
\bea
\hat c\int \rd \hat\pi(\hat\omega)
\ln \frac{Z(\hat{\omega}+\omega)} {Z(\hat\omega)Z(\omega)}
&=& c \int\rd\pi(\omega')
\left\langle\ln \frac{Z_2(\omega,\omega',K)}{Z(\omega)Z(\omega')}\right\rangle_K + \mu\ ,
\label{statpi}
\\
\hat c\int \rd\pi(\omega) \ln \frac{Z(\hat\omega+\omega)} 
{Z(\hat\omega)Z(\omega)} &=& \sum_{k\ge 1} k p_{\hat c}(k)
\int \{\rd \hat\pi\}_{k-1} 
\ln \frac{Z_{\lambda_\varepsilon}(\hat\omega+\{\hat\omega\}_{k-1})}
{Z(\hat\omega)\prod_{\ell=1}^{k-1} Z(\hat\omega_\ell)} +  \hat \mu\ ,
\label{stathatpi}
\eea
with $\mu$ and $\hat\mu$ Lagrange multipliers to take the normalization of $\pi$
and $\hat\pi$ into  account.

The conditions that (\ref{statpi}) must hold for all $\omega$ and
similarly that (\ref{stathatpi}) must hold for all $\hat\omega$ 
can be translated \cite{MezPar01} into
\bea
\hat\pi(\hat{\omega})&=& \frac{c}{\hat c}\int\rd\pi(\omega')~ 
\left\langle\delta(\hat\omega - \hat\Omega(\omega',K))\right\rangle_K\ ,
\label{hatpi}
\\
\pi(\omega)&=&  \sum_{k\ge 1} \frac{k}{\hat c} p_{\hat c}(k)
\int \{\rd \hat\pi\}_{k-1}~\delta\left(\omega - \Omega(\{\hat\omega\}_{k-1})\right)\ ,
\label{pi}
\eea
in which $\hat\Omega(\omega',K)$  and  $\Omega (\{\hat\omega\}_{k-1})$ are defined via
\be
Z(\omega+\hat\Omega(\omega',K))=\frac{Z_2(\omega,\omega',K)}{Z(\omega')}
\qquad \Leftrightarrow \qquad
\hat\Omega(\omega',K)=\frac{K^2}{\omega'}\ ,
\label{defhatOm}
\ee
and 
\be
\Omega(\{\hat\omega\}_{k-1})= i\lambda_\varepsilon+ \sum_{\ell=1}^{k-1} \hat\omega_\ell\ ,
\label{defOm}
\ee
respectively. Given that $\pi$ is normalized, it follows from (\ref{hatpi}) 
that the same is true for $\hat\pi$, provided $\hat c= c$, so the fixed point
equations take their final form as
\bea
\hat\pi(\hat{\omega})&=& \int\rd\pi(\omega')~ 
\left\langle\delta(\hat\omega - \hat\Omega(\omega',K))\right\rangle_K\ ,
\label{hatpif}
\\
\pi(\omega)&=&  \sum_{k\ge 1} \frac{k}{c} p_{c}(k)
\int \{\rd \hat\pi\}_{k-1}~\delta\left(\omega - \Omega(\{\hat\omega\}_{k-1})\right)\ .
\label{pif}
\eea
These equations can be seen as special cases of the general framework derived 
in \cite{Ku+07}, when restricted to harmonically coupled random systems. In \cite{Ku+07} 
it is shown that they hold --- unmodified --- for non-Poissonian degree distributions 
as well, as long as the average connectivity in these systems remains finite. 

Note that for all $\varepsilon >0$,  $\pi$ and $\hat\pi$  --- self-consistently 
--- have support in ${\rm Re}~\omega >0$ and ${\rm Re}~\hat \omega >0$ as required. 
The equations take a form that suggests solving them via a stochastic population-based 
algorithm, as described in Appendix A.

For the thermodynamic limit of the spectral density we obtain from (\ref{Zn}), 
(\ref{avlog}) and (\ref{ZnGi})-(\ref{defZ2s}) that
\bea
\overline{\rho(\lambda)} &=& 
\frac{1}{\pi}{\rm Im}\, \sum_{k=0}^\infty p_c(k) \int \{\rd \hat\pi\}_k ~
\frac{i}{i \lambda_\varepsilon + \{\hat\omega\}_k}\nonumber\\
&=& \frac{1}{\pi} \sum_{k=0}^\infty p_c(k) \int \{\rd
\hat\pi\}_k  ~ \frac{{\rm Re}\big(\{\hat\omega\}_k +\varepsilon\big)}
{\big({\rm Re}(\{\hat\omega\}_k +\varepsilon)\big)^2+ \big(\lambda + {\rm Im}\,
\{\hat\omega\}_k\big)^2}\ .
\label{spec1}
\eea
This expression has a natural interpretation as a sum of contributions of
local-densities of state of sites with connectivities $k$, weighted according
to their probability of occurrence. Referring to (\ref{avlog}), we may further
identify the
\be
\sigma_k^2 = \frac{1}{\pi} {\rm Im}~\frac{i}{i \lambda_\varepsilon + \{\hat\omega\}_k}
\ee
as realizations of the variance of (Gaussian) marginals on sites of coordination
$k$.

With an eye towards disentangling singular (pure point) and continuous contributions
to the spectral density, we find it useful to define 
\be
P(a,b)= \sum_k p_c(k) \int \{\rd \hat\pi\}_k ~\delta\left(a - 
{\rm Re}~ \{\hat \omega\}_k \right) \delta\left(b - {\rm Im}~ \{\hat \omega\}_k\right)\ ,
\ee
with $a \ge 0$ by construction. The density of states can then be expressed as
an integral over $P$,
\be
\overline{\rho(\lambda)}= \int \frac{\rd a ~ \rd b}{\pi}  ~ P(a,b) ~ 
\frac{a+\varepsilon}{(a+\varepsilon)^2+ (b+\lambda)^2}\ .
\label{specab}
\ee
Noting the singlular nature of the above integrand in the limit $\varepsilon \to 0$ 
for $a=0$, we propose to isolate possible singular contributions to the spectral 
density by writing
\be
P(a,b)= P_0(b) \delta(a) + \tilde P(a,b)\ .
\ee
This gives
\be
\overline{\rho(\lambda)} =  \int \rd b ~ P_0(b) {\cal L}_\varepsilon(b+\lambda) 
+  \int_{a>0} \frac{\rd a ~ \rd b}{\pi}~ \tilde P(a,b) ~ 
\frac{a+\varepsilon}{(a+\varepsilon)^2+ (b+\lambda)^2}\ ,
\label{specab2}
\ee
in which ${\cal L}_\varepsilon$ denotes a Lorentzian of width $\varepsilon$.
Our results below strongly suggest that, when the limit $\varepsilon \to 0$ is taken ---
thereby ${\cal L}_\varepsilon(x) \to \delta(x)$ --- a non-zero value of
\be
P_0(-\lambda) = \lim_{\varepsilon \to 0} \int \rd b ~ P_0(b) {\cal L}_\varepsilon(b+\lambda)
\ee
gives the contribution of the pure-point spectrum, originating
from localized states, to the overall spectral density.

This concludes the general framework.
%----------------------------------------------------------------------------------
\section{Results}
%----------------------------------------------------------------------------------

In what follows, we report results for a variety of different ensembles of
sparse random matrices, in order to explore the capabilities and limitations
of our approach. In order to properly appreciate the results presented below,
it is worth pointing out that within our stochastic population-dynamics based
approach to solving the fixed point equations (\ref{hatpif}) and (\ref{pif}),
the integrals (\ref{spec1}), or (\ref{specab}), (\ref{specab2}) are evaluated
by {\em sampling\/} from a population. Denoting by ${\cal N}$ the number of
samples $(a_i,b_i)$ taken, we have, e.g.,
\be
\overline{\rho(\lambda)} \simeq  \frac{1}{{\cal N}}
\left[ \sum_{{i=1\atop a_i=0}}^{{\cal N}} {\cal L}_\varepsilon(b_i+\lambda) 
+  \frac{1}{\pi} \sum_{{i=1 \atop a_i>0}}^{{\cal N}}~
\frac{a_i+\varepsilon}{(a_i+\varepsilon)^2+ (b_i+\lambda)^2}\right]
\label{specsamp0}
\ee
as an approximation of (\ref{specab2}). The $\varepsilon \to 0$-limit is clearly
singular in the first contribution to (\ref{specsamp0}). If $b_i+\lambda \ne 0$ for
all $b_i$ in the sample, one obtains zero in the $\varepsilon \to 0$-limit, whereas
one obtains a diverging contribution, if $b_i+\lambda = 0$ for at least one $b_i$
in the sample. The second alternative will quite generally be an event of 
probability zero, so a small regularizing $\varepsilon >0$ must be kept in order
to `see' this contributions (if it exists). In what follows, we shall refer to
the two contributions to (\ref{specab2}), as $\overline{\rho_s(\lambda)}$ and 
$\overline{\rho_c(\lambda)}$, with
\be
\overline{\rho_s(\lambda)} \simeq  \frac{1}{{\cal N}}
\sum_{{i=1\atop a_i=0}}^{{\cal N}} {\cal L}_\varepsilon(b_i+\lambda) \ , \quad
\overline{\rho_c(\lambda)} \simeq \frac{1}{\pi{\cal N}} \sum_{{i=1 \atop a_i>0}}^{{\cal N}}~
\frac{a_i+\varepsilon}{(a_i+\varepsilon)^2+ (b_i+\lambda)^2}\ .
\label{specsamp}
\ee
The population-dynamics algorithm itself is run with a small regularizing 
$\varepsilon >0$ (as required in (\ref{Zn}) to guarantee existence of the integral).
While running the algorithm, we use $\varepsilon = 10^{-300}$, which is close to the 
smallest representable real number in double-precision arithmetic on the machines
used for the numerics. 

%----------------------------------------------------------------------------------
\subsection{Poisson Random Graphs --- Gaussian Couplings}
%----------------------------------------------------------------------------------

\begin{figure}[hb]
\epsfig{file=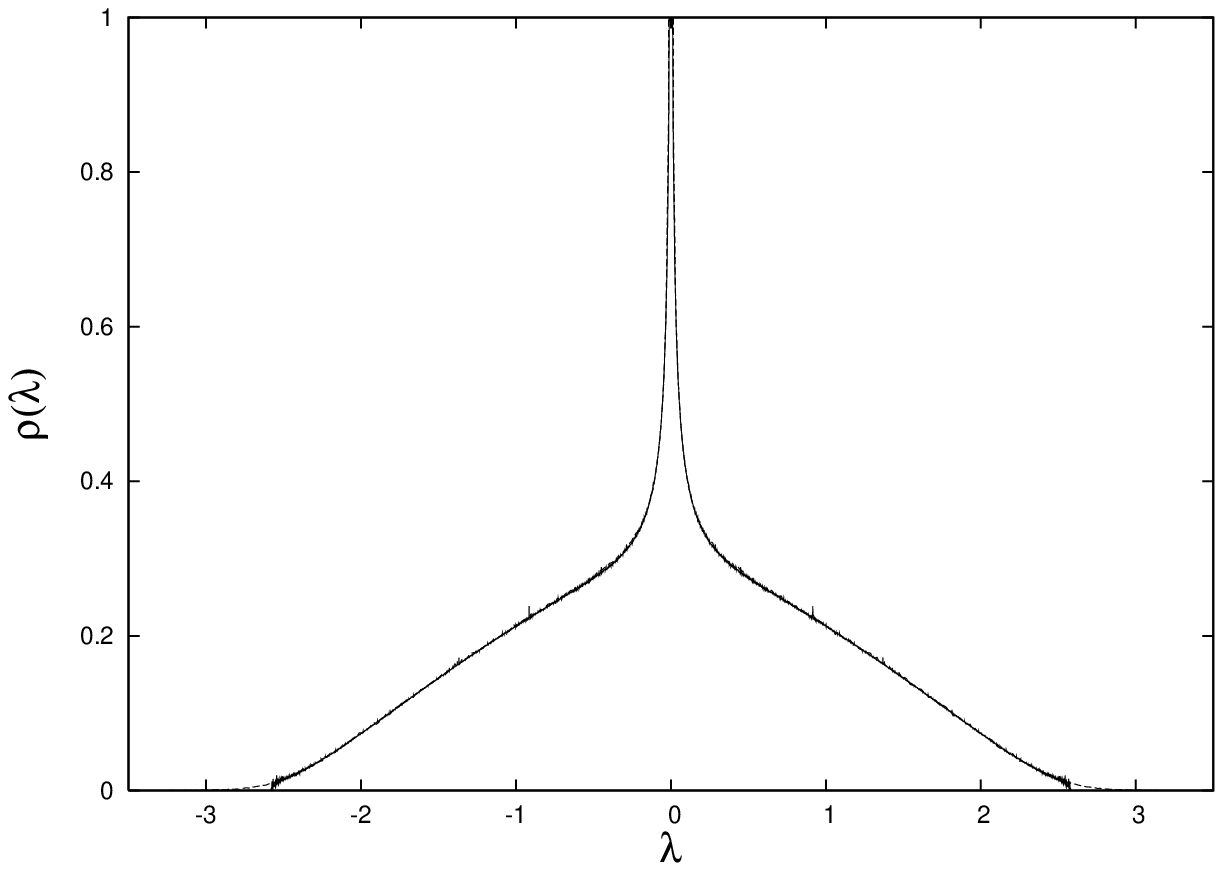,width=0.49\textwidth}\hfill
\epsfig{file=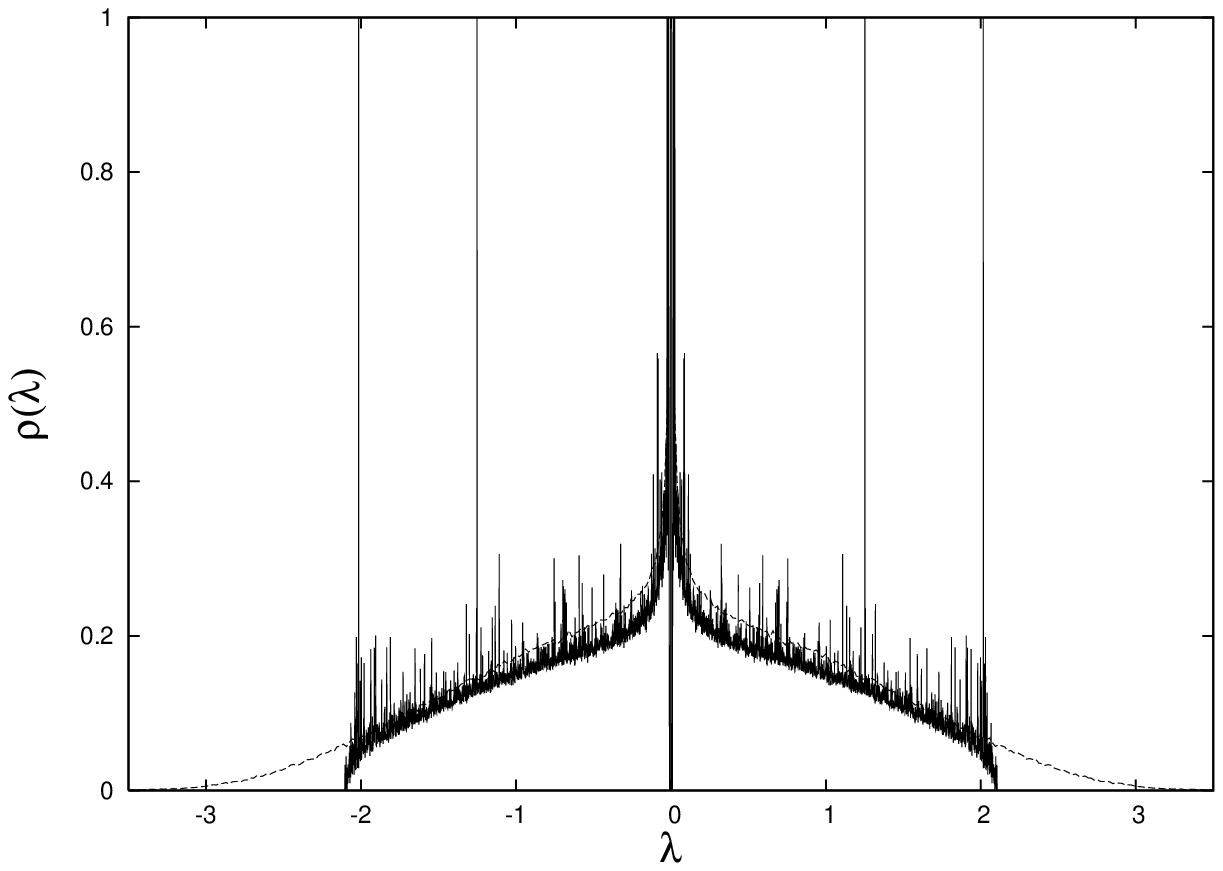,width=0.49\textwidth}
\caption{Spectral density for matrices defined on Poissonian random graphs with 
$c=4$ (left panel) and $c=2$ (right panel), having Gaussian random couplings 
with $\langle K_{ij}^2\rangle=1/c$. Full line: results obtained from the present theory; 
dashed line: results obtained from a sample of 2000$\times$2000 matrices. In both
cases $\varepsilon = 10^{-300}$  was used in the evaluation of (\ref{specsamp0}).}
\label{fig:p4.2.g}
\end{figure}

Our first results pertain to sparse matrices defined on Poisson random graphs, with
Gaussian couplings. The left panel of Fig. \ref{fig:p4.2.g} shows spectral densities 
for the case of mean connectivity $c=4$, having Gaussian random couplings with $\langle
K_{ij}^2\rangle=1/c$. For this system we find an integrable power-law 
divergence of the form
\be
\overline \rho (\lambda) \simeq 0.05|\lambda|^{-0.61}\ ,\qquad \lambda \to 0\ ,
\ee
and a $\delta$ peak at $\lambda=0$, the latter originating from isolated sites in
the ensemble. Results of numerical diagonalizations (using a sample of $500$  
$N \times N$ matrices with $N=2000$ are shown for comparison, and the agreement 
is excellent.

\begin{figure}
\epsfig{file=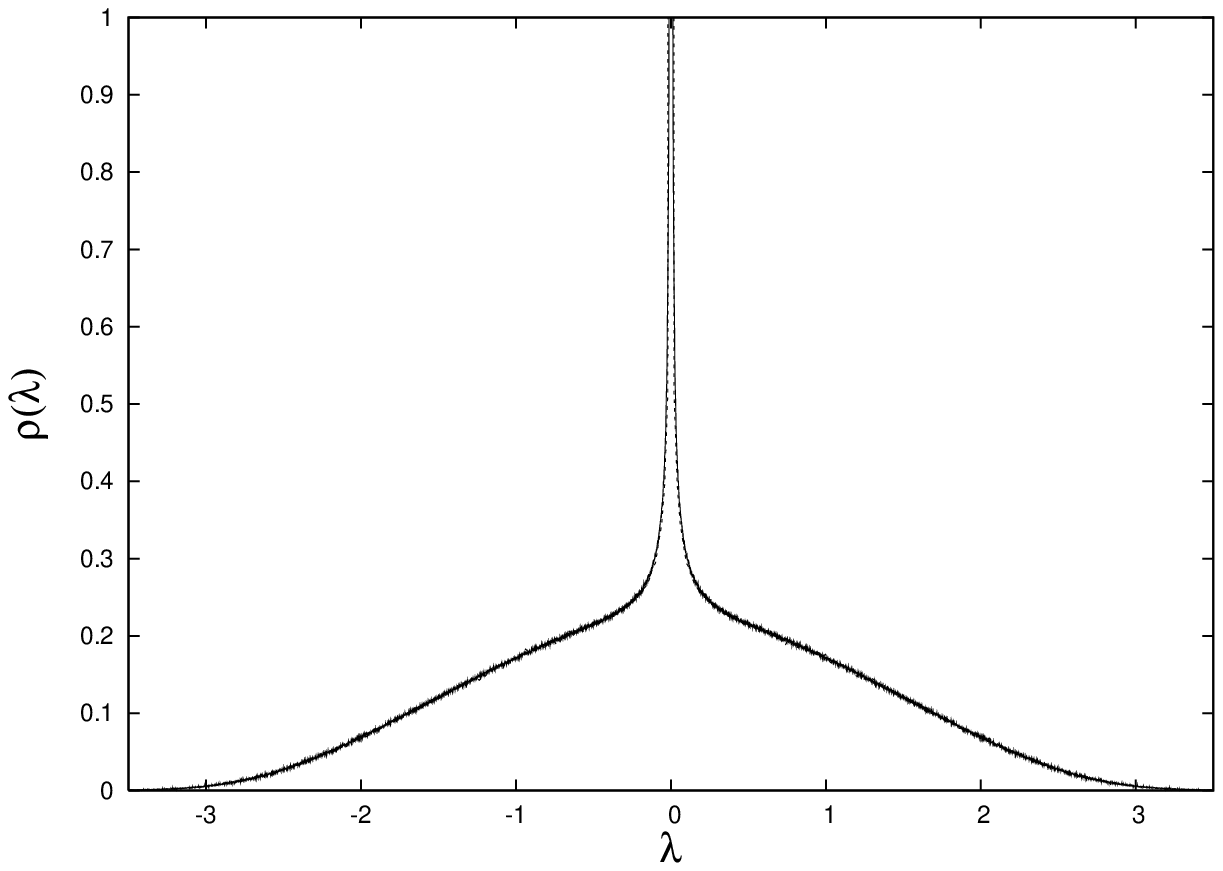,width=0.49\textwidth}\hfill
\epsfig{file=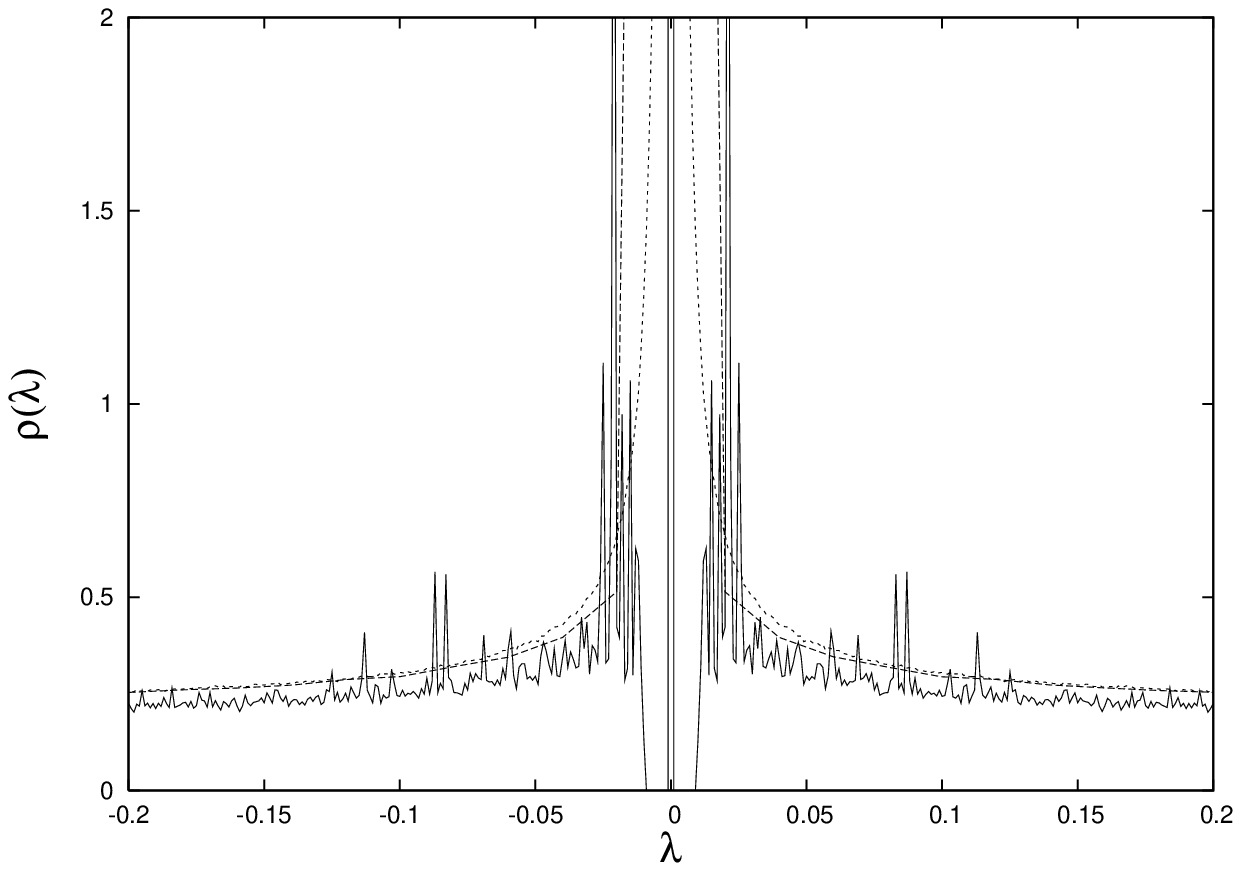,width=0.49\textwidth}
\centering{\epsfig{file=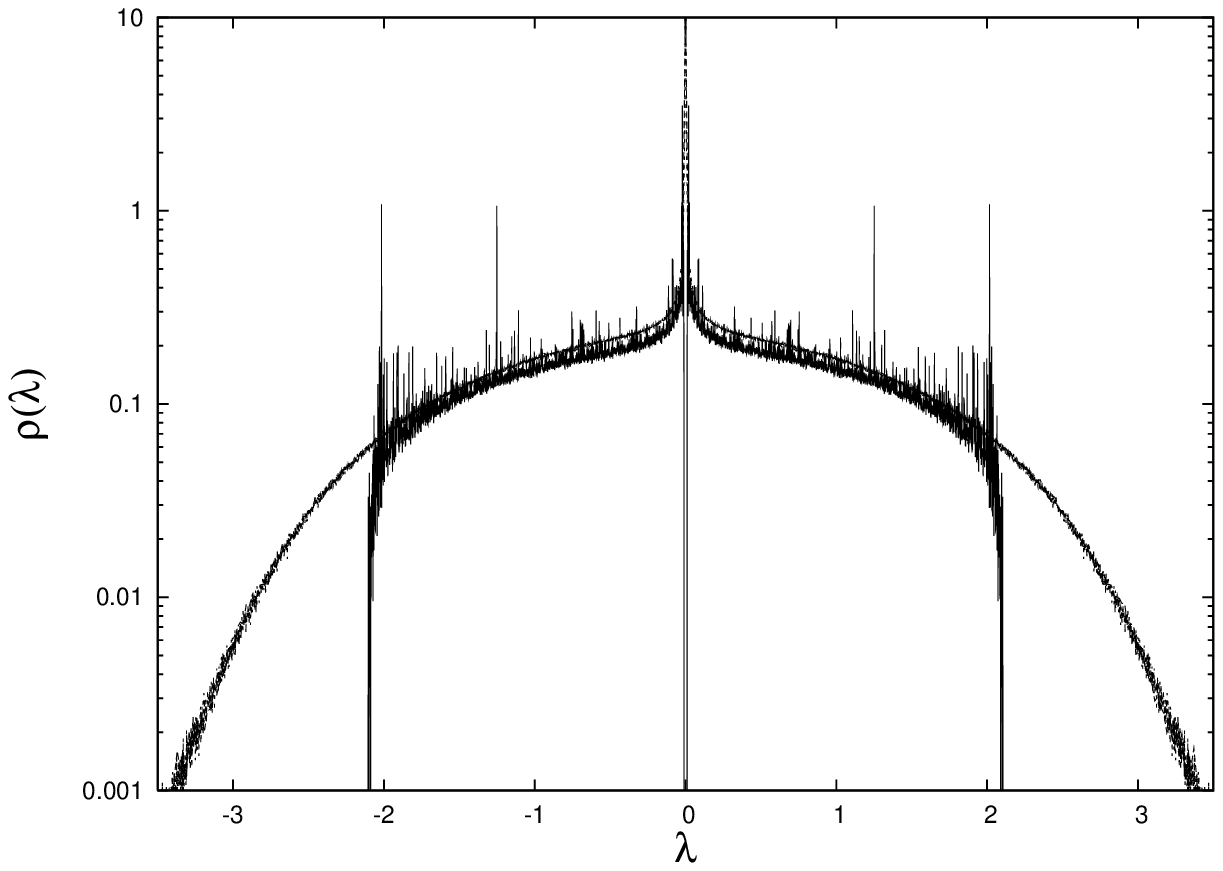,width=0.49\textwidth}}
\caption{Upper left panel: Spectral density for matrices defined on Poissonian 
random graphs with $c=2$ as in the previous figure, but now evaluated with a 
regularizing $\varepsilon = 10^{-3}$  in (\ref{specsamp0}) (full line). At the 
resolution given the result is indistinguishable from the numerical simulation 
results (dashed line). Upper right panel: zoom into the central region comparing
results obtained with the small regularizer, exhibiting a gap around the central 
peak (full line), with a larger regularizer $\varepsilon = 10^{-3}$ (short dashed 
line) and with results of numerical diagonalization (long dashed line). The same 
comparison is made in the lower panel for a larger portion of the spectrum on a 
logarithmic scale. The regularized $\varepsilon = 10^{-3}$-results are  on this 
scale indistinguishable from those of the numerical simulations. Note the 
localization transition and the Lifshitz tails as discussed in the main text.}
\label{fig:p2.g}
\end{figure}

The behaviour changes rather drastically if the average connectivity is reduced to 
$c=2$ --- a value closer to the percolation threshold $c_c=1$. In this case the 
spectral density shows strong fluctuations, when evaluated with the same small 
regularizer. These originate from $\overline{\rho_s}$ in (\ref{specsamp}), 
and are related to the pure point spectrum associated with localized eigenstates 
coming from a collection of isolated finite clusters of all sizes in the ensemble.
These exist for $c=4$ as well, but their contribution is too small to be easily
notable when combined with $\overline{\rho_c}$ in (\ref{specsamp0}). In addition,  
there is a central $\delta$ peak as in the $c=4$-case which appears to be separated 
from the main bands by a gap; see the second panel in Fig  \ref{fig:p2.g}. The 
agreement with results of numerical diagonalization is fairly poor as it stands;
in particular, exponential tails of localized states extending beyond the apparent 
edge of the central band are missed in this way. However, when (\ref{specsamp0}) 
is evaluated with a regularizing $\varepsilon = 10^{-3}$ comparable to the 
resolution of the $\lambda$-scan, the agreement is once more excellent as shown 
in Fig \ref{fig:p2.g}. It is worth noting in this context that numerical simulations, 
in which binning of eigenvalues is used to determine the the spectral density {\em 
also imply a form of regularization}, and they do not distinguish continuous and 
singular contributions to the DOS if the distribution of the singular contributions 
is itself reasonably uniform. 

When displayed on a logarithmic scale, the results clearly reveal two interesting 
features: (i) a localization transition at $\lambda_c \simeq 2.295$, characterised 
by a vanishing continuous contribution $\overline{\rho_c}$ to (\ref{specsamp0})
for $|\lambda| > \lambda_c$, and (ii) exponential (Lifshitz) tails  \cite{Khor+06} 
in the spectral density, related to localized states represented by the singular 
contribution $\overline{\rho_s}$ to (\ref{specsamp0})), and exhibited only 
through regularization. We shall substantiate this analysis in the following 
sub-section by looking at the behaviour inverse participation ratios. The same 
phenomena are seen for $c=4$, where $\lambda_c \simeq 2.581$.

\begin{figure}[ht]
\epsfig{file=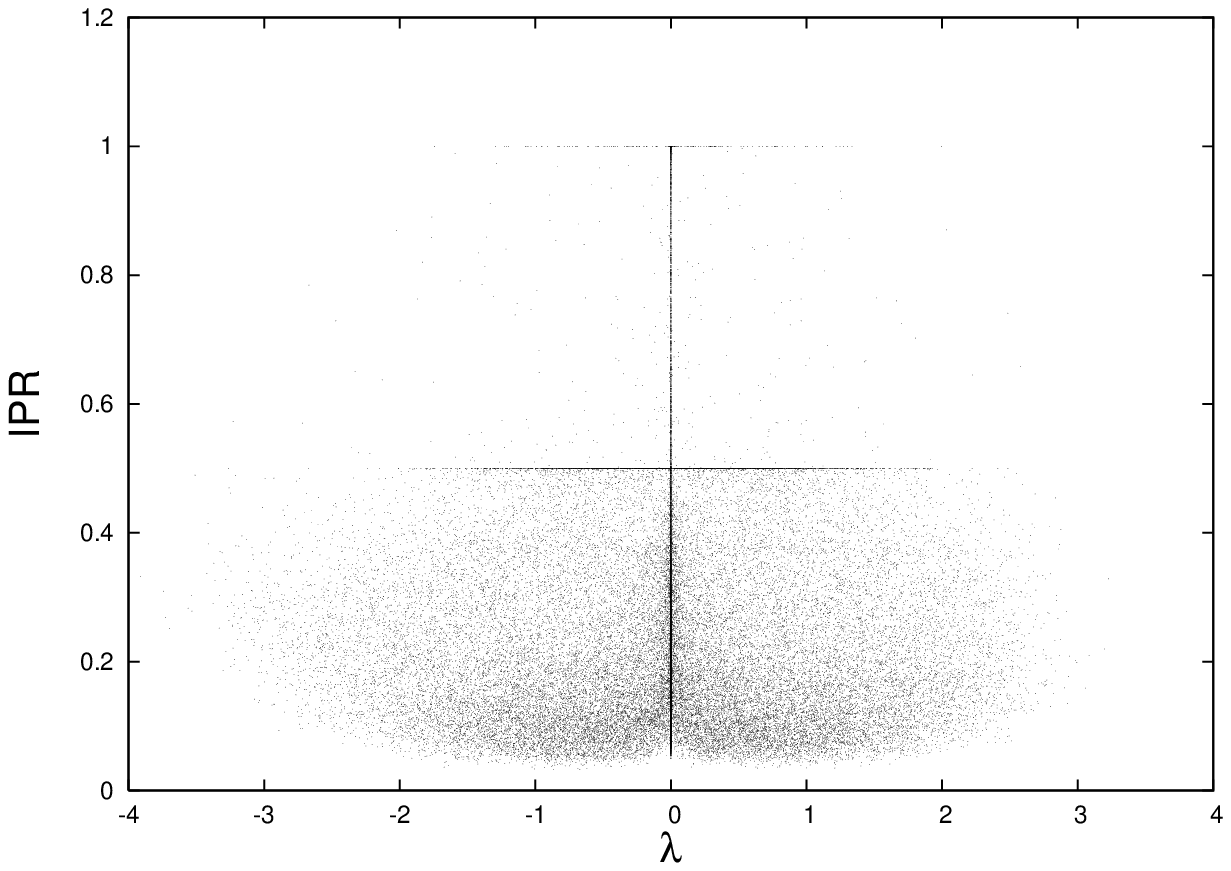,width=0.49\textwidth}\hfill
\epsfig{file=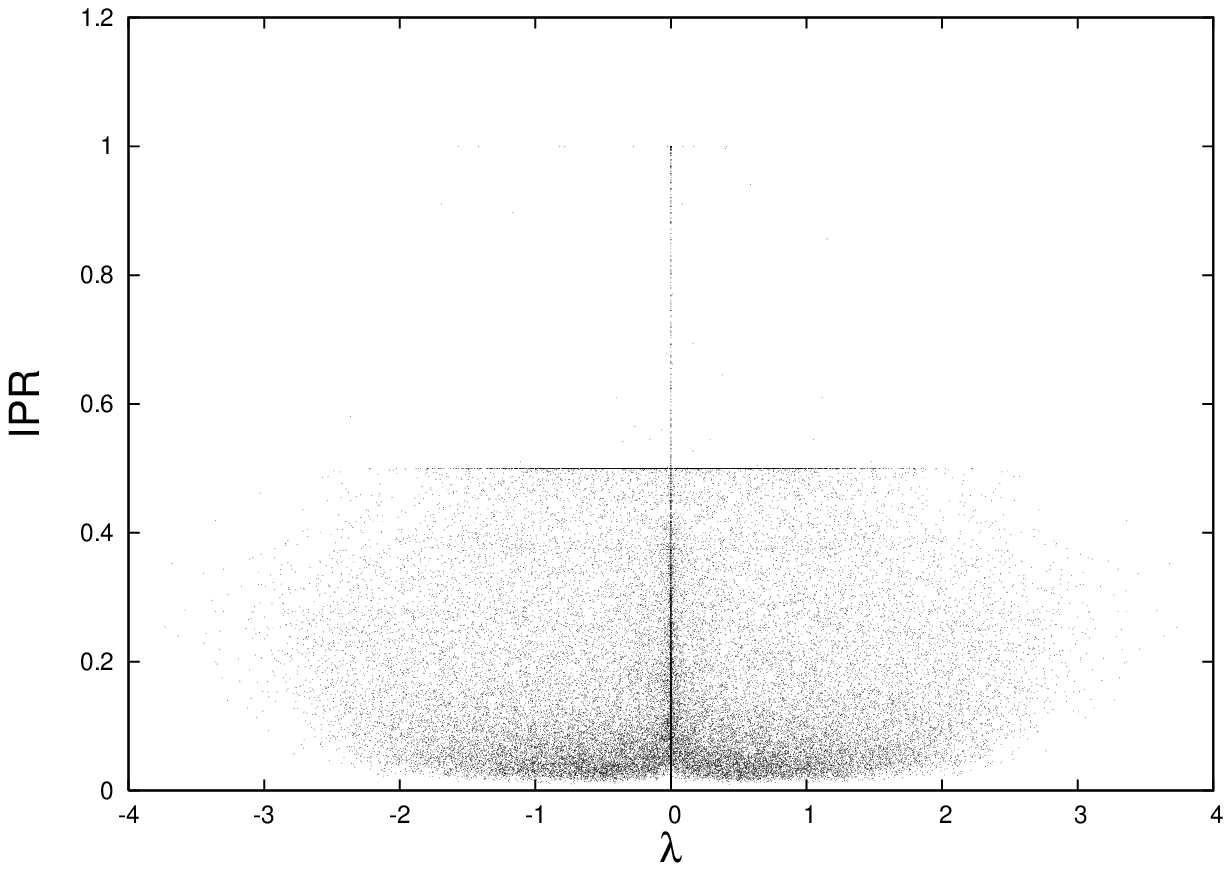,width=0.49\textwidth}\\
\epsfig{file=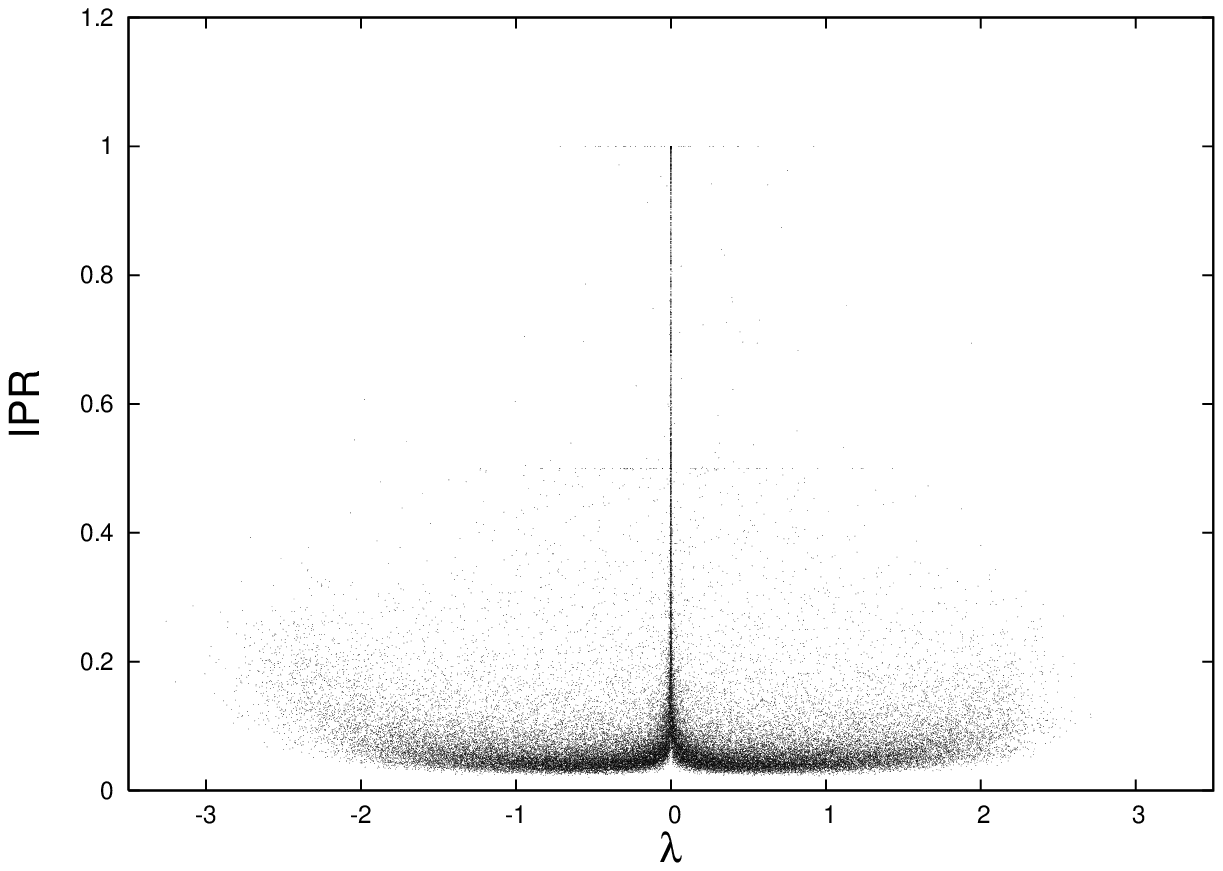,width=0.49\textwidth}\hfill
\epsfig{file=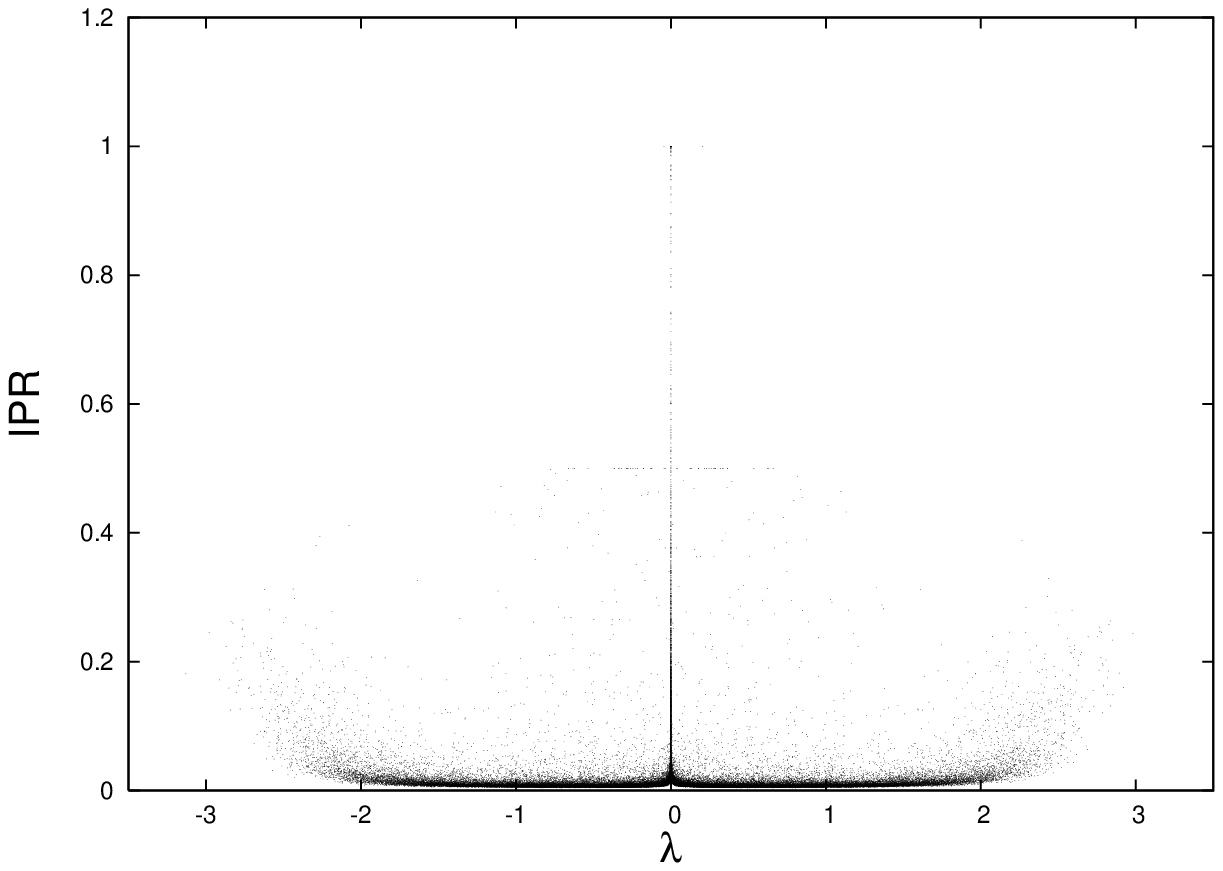,width=0.49\textwidth}
\caption{Scatterplots showing eigenvalue against IPRs for Poissonian random graphs 
with $c=2$ (first row) and $c=4$  (second row). The graphs in the left column correspond
to $N=100$, those in the right column to $N=1000$. }
\label{fig:IPR}
\end{figure}
%----------------------------------------------------------------------------------
\subsection{Inverse Participation Ratios and Localization}
%----------------------------------------------------------------------------------

In order to substantiate our identification of singular and continuous contributions
to the spectral densities we look at Inverse Participation Ratios (IPRs) of eigenstates 
as obtained from numerical diagonalizations. Given eigenvectors $\bm v$ of a (random) 
matrix, their IPRs are defined as
\be
{\rm IPR}(\bm v) = \frac{\sum_{i=1}^N v_i^4}{\Big(\sum_{i=1}^N v_i^2\Big)^2}\ .
\ee
As eigenvectors can always be chosen to be normalized, we see that IPRs remain of
order 1 for localized states which have a few $\cO(1)$ eigenvector components ---
the extreme case being ${\rm IPR}(\bm v)=1$ for $v_i = \delta_{i,i_0}$ --- whereas they 
are $\cO(N^{-1})$ for fully extended states for which $v_i = \cO(N^{-1/2})$ for 
all $i$.

Here we only produce a qualitative comparison for the two cases studied in the previous
subsection, comparing IPRs computed for systems of size $N=100$ and $N=1000$, and using
scatter-plots of IPRs vs eigenvalues to exhibit the salient features. As clearly visible,
there remains a substantial fraction of states at all $\lambda$ in the $c=2$ case, which
do {\em not\/} exhibit the $N^{-1}$ scaling of IPRs expected for delocalized states; 
the tails, and a small central band in particular appear to be {\em dominated\/} by 
localized states. By contrast in the $c=4$ case there is a notable depletion of states
with $\cO(1)$ IPRs, except for $\lambda=0$ and in the tails of the spectrum. These
findings are entirely consistent with our identifications made in the previous 
subsection. We note that the role of regularization in identifying localized states 
has been pointed out before using heuristics related to the evaluation of {\em local\/} 
densities of state \cite{Cil+05}. 

We shall return to this issue in greater quantitative detail in a separate paper
devoted to Anderson localization in discrete random Schr\"odinger operators defined
on sparse random graphs \cite{KuMour08}.

\begin{figure}[ht]
\hfil\epsfig{file=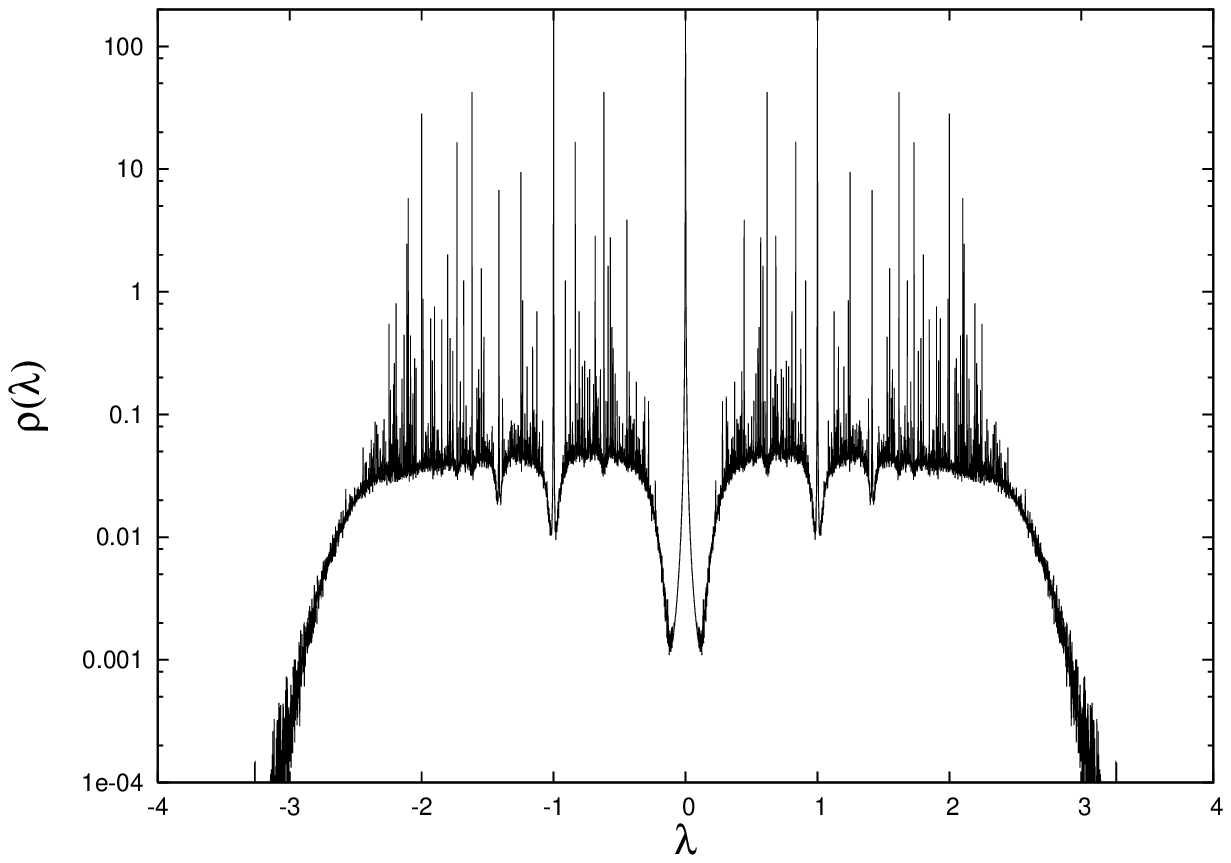,width=0.75\textwidth}\hfill\\
\epsfig{file=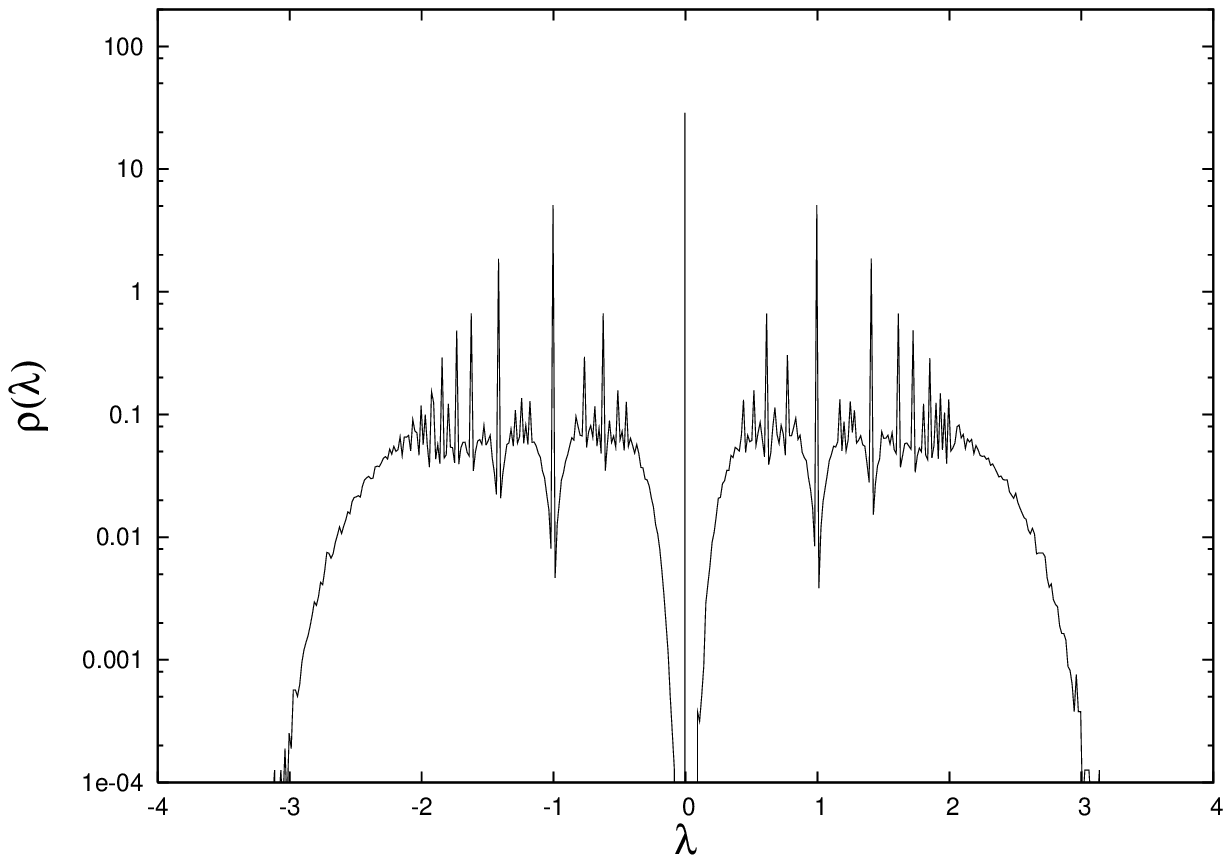,width=0.49\textwidth}\hfill
\epsfig{file=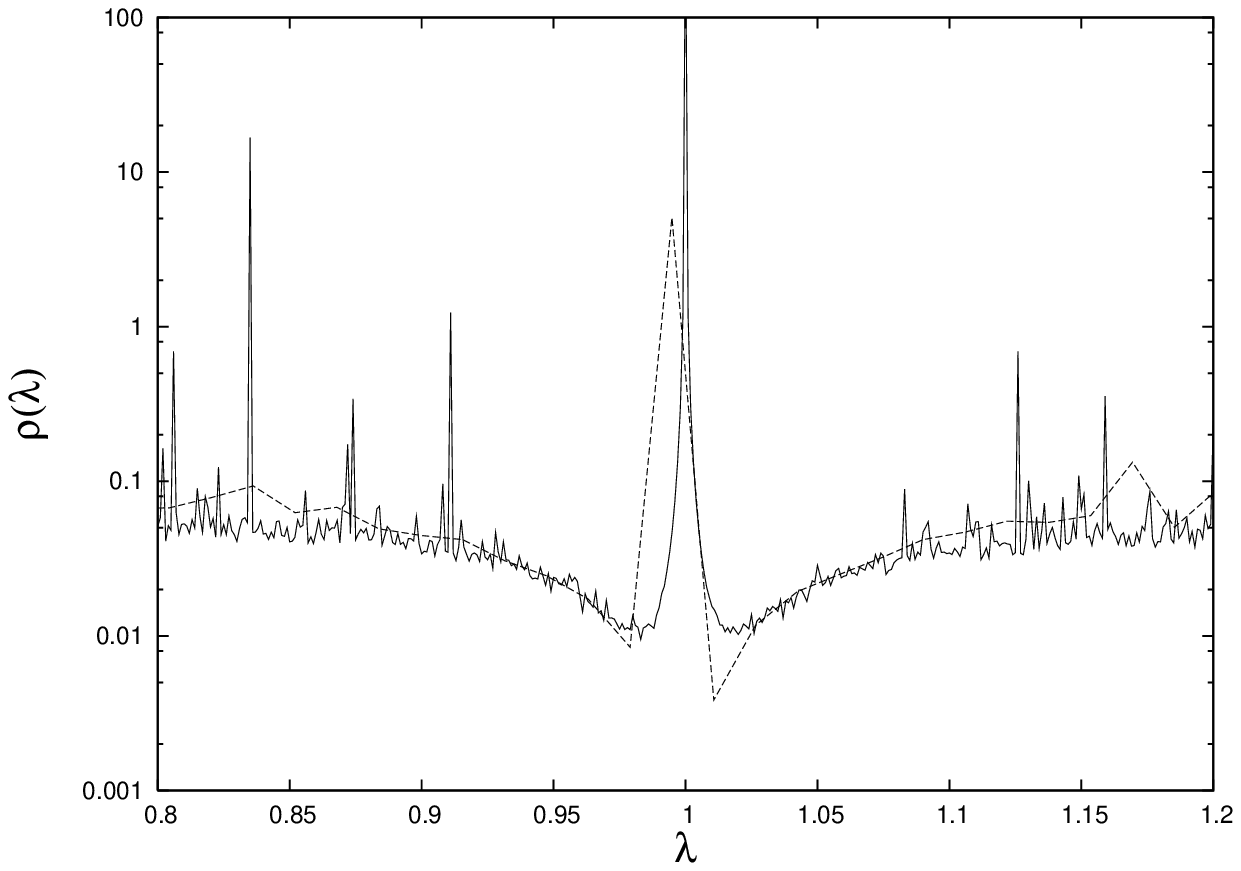,width=0.49\textwidth}
\caption{Comparison of spectral density for $K_{ij}= \pm 1/\sqrt c$, on a Poissonian 
random graph with $c=1$ as computed via the present algorithm (main panel) with results 
from numerical diagonalisation of $N\times N$ matrices of the same type with $N=2000$
(lower left) and a direct comparison in the region around $\lambda=1$.}
\label{fig:pb1}
\end{figure}
%----------------------------------------------------------------------------------
\subsection{Poisson Random Graphs --- Bimodal Couplings}
%----------------------------------------------------------------------------------

We can also look at coupling distributions different from Gaussian for the non-zero
couplings, e.g. fixed $K_{ij}=1/\sqrt c$ or bi-modal $K_{ij}=\pm 1/\sqrt c$. As noted 
before \cite{RodgBray88}, both give rise to the same spectral densities on large 
sparse (tree-like) graphs due to the absence of frustrated loops. It can also be 
seen as a consequence of the appearance of $K^2$ in (\ref{defhatOm}).

We choose a Poissonian random graph {\em at\/} the percolation threshold $c=1$ as
an example that allows us to highlight both the strengths and the limitations of the
present approach. It is known that all states will be localized for this system. 
In Fig \ref{fig:pb1} we compare results of a $\lambda$-scan with resolution $\delta 
\lambda=10^{-3}$, using a regularizer $\varepsilon=10^{-4}$ for the scan. The smaller
panels exhibit numerical diagonalization results, as well as a comparison between the
two using a zoom into the region around $\lambda=1$.

On the side of the strengths, we note that the spectral density obtained from our
algorithm is able to display more details than can be exposed by simulation results
obtainable at reasonable effort. On the downside, one might note that the results
for this system attain the status of semi-quantitative results, as they do depend 
on the chosen regularization, though in fairness it should be said that the same 
applies to the results obtained via numerical diagonalization where results vary 
with the binning resolution. In the present case this is due to the fact that the
spectrum for most parts consists of a dense collection of $\delta$ peaks \cite{Goli03}.
A notable deficiency is the broadening of delta-peaks into Lorentzians of finite width, 
which creates artefacts around isolated delta-peaks, exemplified here by the peak at 
$\lambda=0$. Since the origin of this deficiency is understood, more precise details 
can, if desired, be recovered by choosing a smaller regularizing $\varepsilon$.

%----------------------------------------------------------------------------------
\subsection{Regular and Scale-Free Random Graphs}
%----------------------------------------------------------------------------------

In the present section we consider matrices defined on regular and scale-free 
random graphs.

%----------------------------------------------------------------------------------
\subsubsection{Regular Random Graphs}
%----------------------------------------------------------------------------------

\begin{figure}[ht]
\epsfig{file=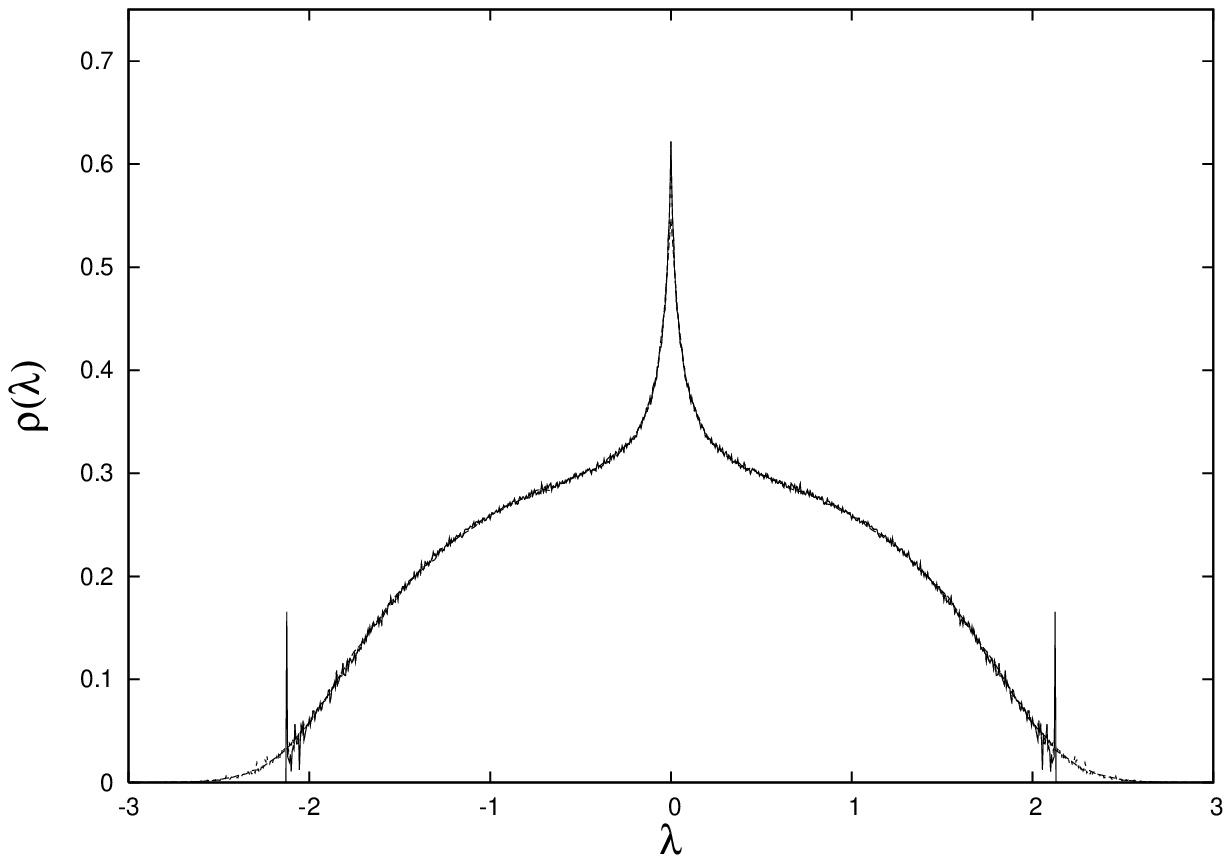,width=0.49\textwidth}\hfill
\epsfig{file=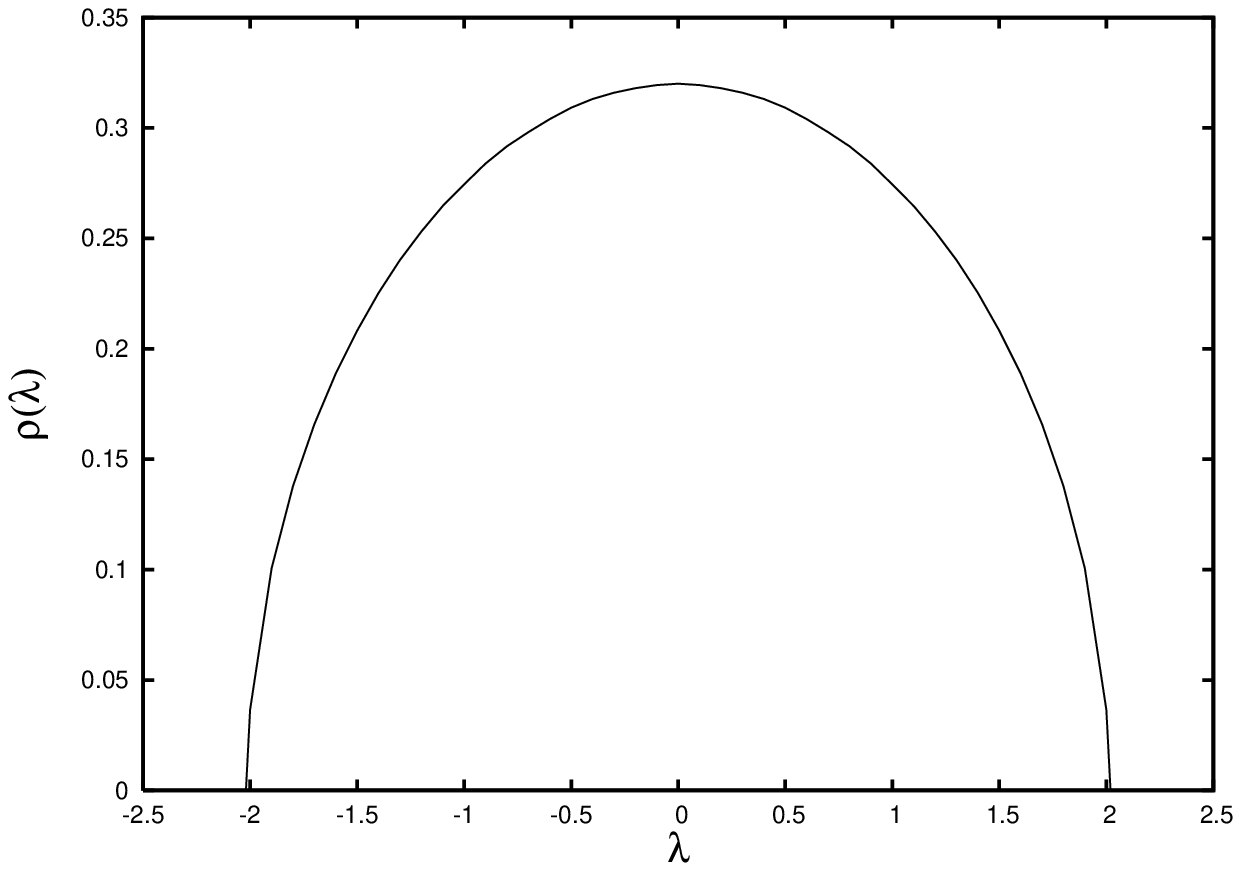,width=0.49\textwidth}
\caption{Spectral densities for  a  random graph with fixed connectivity $c=4$ (left), 
and  on a random graph with fixed non-random connectivity $c=100$ (right).}
\label{fig:fixk}
\end{figure}

Our theory applies unmodified to matrices defined on graphs with degree distributions
other than Poissonian, as long as the {\em mean connectivity remains finite}. We use 
this fact to obtain spectra of matrices with Gaussian random couplings defined on 
regular random graphs with fixed connectivity $c$, choosing $\langle K_{ij}^2\rangle
=1/c$ for the couplings. Results for $c=4$ and $c=100$ are shown in Fig. \ref{fig:fixk}. 
The $c=4$ results are compared with simulations, with results analogous to previous 
cases, including the presence of a localization transition at $\lambda_c \simeq 2.14$

The second example is chosen as a test to see the semicircular law \cite{Wign58} 
reemerge in the limit of large (though finite) connectivity. This limit can also 
be extracted from the fixed point equations. It is somewhat easier to verify for
results pertaining to single instances \cite{Rog+08} than for the ensemble.

%----------------------------------------------------------------------------------
\subsubsection{Scale-Free Graphs}
%----------------------------------------------------------------------------------

\begin{figure}[ht]
\epsfig{file=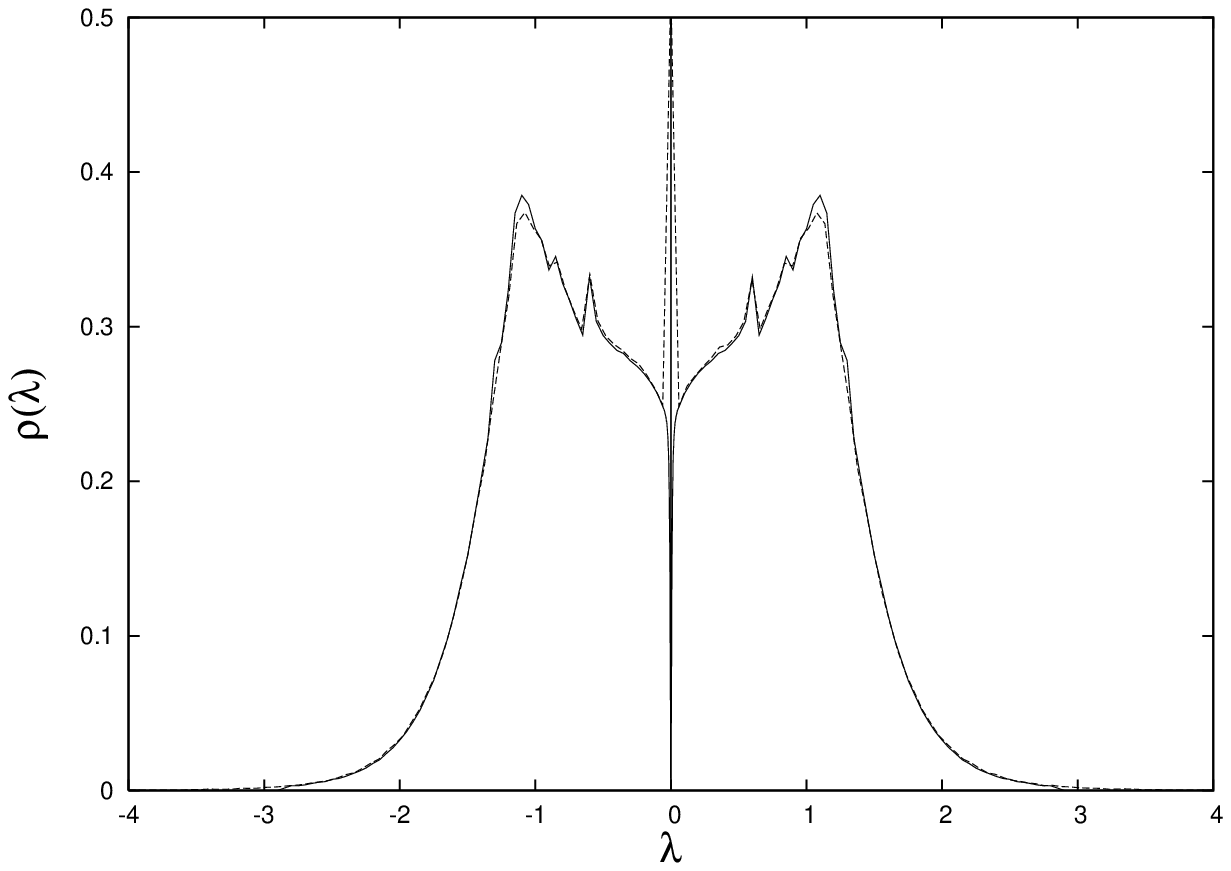,width=0.49\textwidth}\hfill
\epsfig{file=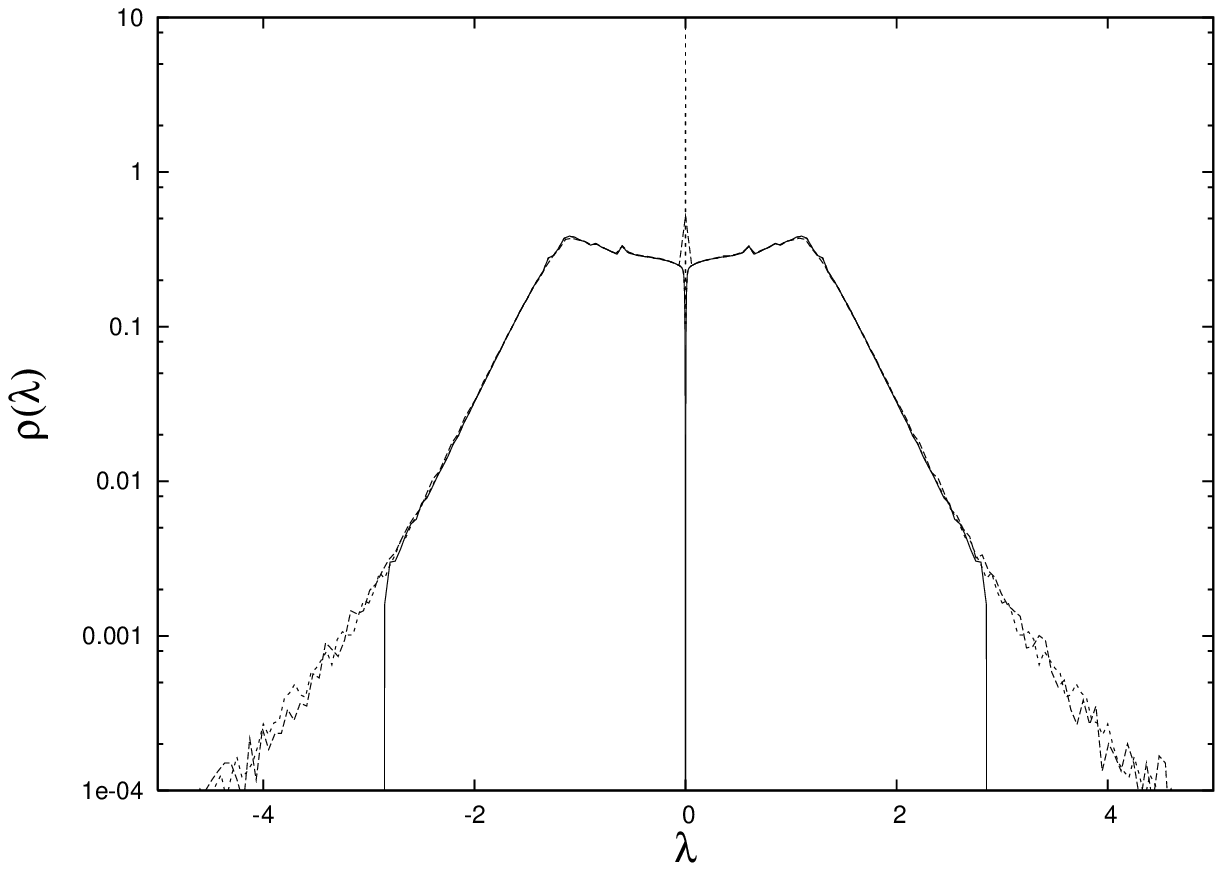,width=0.49\textwidth}
\caption{Spectral density for for $K_{ij}= \pm 1/\sqrt c$ on a random graph with 
power-law degree distribution of average connectivity $c\simeq 2.623$. 
Left panel: results obtained with small regularizer (full line), and numerical 
diagonalization results from a sample of 500 matrices of dimension $N=2000$ (dashed 
line). Right panel: the same results displayed on a logartithmic scale, this time with
results regularized at $\varepsilon = 10^{-3}$ (short dashed line) included.}
\label{fig:pow24}
\end{figure}

We have also looked at a scale free graph with connectivity distribution given by
$p(k) = P_0 k^{-\gamma}$ with $\gamma=4$ and a lower cut-off at $k=2$. Results shown 
in Fig. \ref{fig:pow24} reveal a continuous central band, and localized states for 
$|\lambda| > \lambda_c \simeq 2.85$ much as in the other cases. For the present 
system, the tails in the spectral density follow a power law of the form 
$\rho(\lambda) \sim \lambda^{1- 2\gamma}$  \cite{Dorog+03, MihPap02}. 

Comparison with exact diagonalization results is facilitated by a fast algorithm 
that allows to generate sparse graphs with arbitrary degree distribution 
\cite{MourKab03}.

%----------------------------------------------------------------------------------
\subsection{Graph Laplacians}
%----------------------------------------------------------------------------------

Let us finally look at matrices row-constraints, such as related to discrete
graph-Laplacians.

The discrete graph Laplacian of a graph with connectivity matrix $C= \{c_{ij}\}$
has matrix elements
\be
\Delta_{ij} = c_{ij} - \delta_{ij} \sum_k c_{ik}\ .
\ee
A quadratic form involving the Laplacian can be written in the form
\be
\frac{1}{2} \sum_{ij} \Delta_{ij} u_i u_j = -\frac{1}{4} \sum_{ij} c_{ij} (u_i-u_j)^2\ .
\ee
As before we shall be interested in more general matrices with zero row-sum 
constraint of the form
\be
M_{ij} = c_{ij}K_{ij} - \delta_{ij} \sum_k c_{ik}K_{ij}\ .
\ee

\begin{figure}[ht]
\epsfig{file=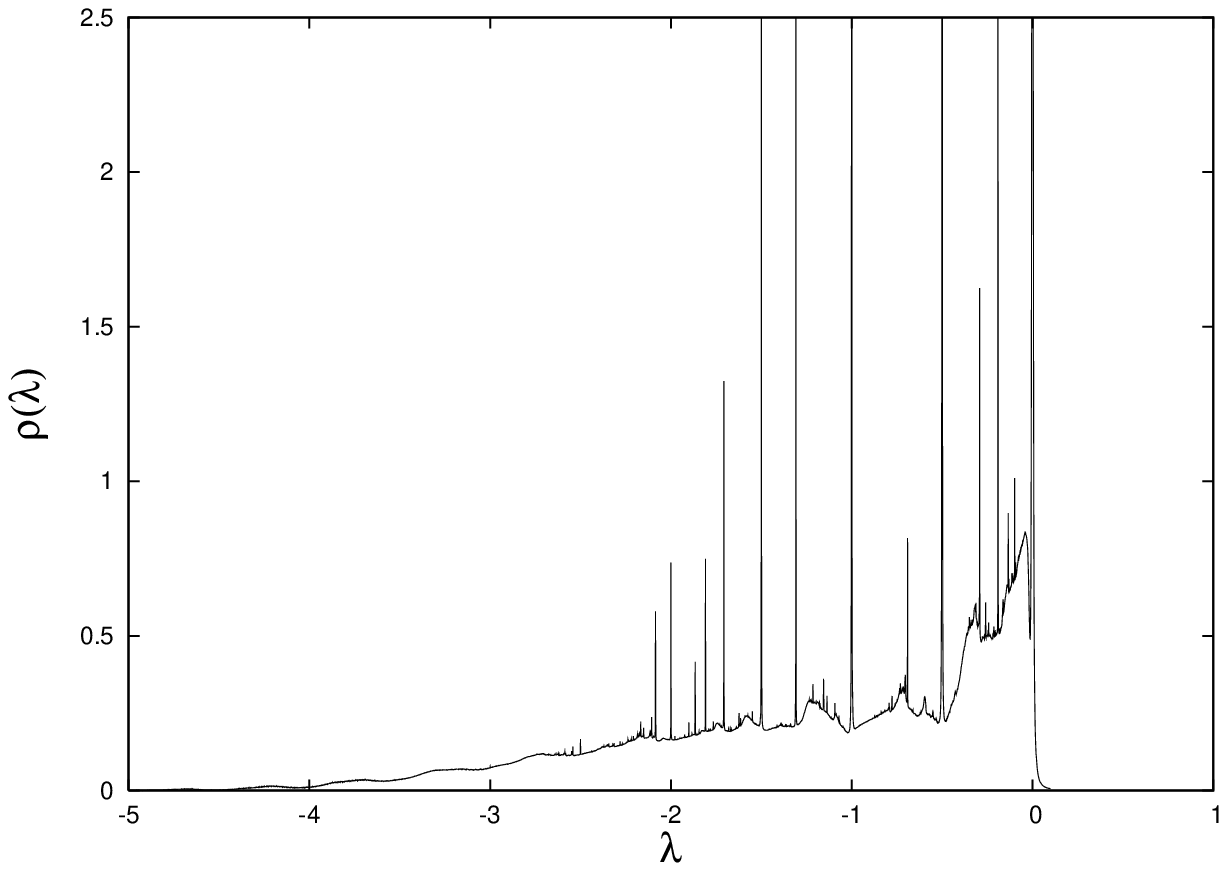,width=0.49\textwidth}\hfill
\epsfig{file=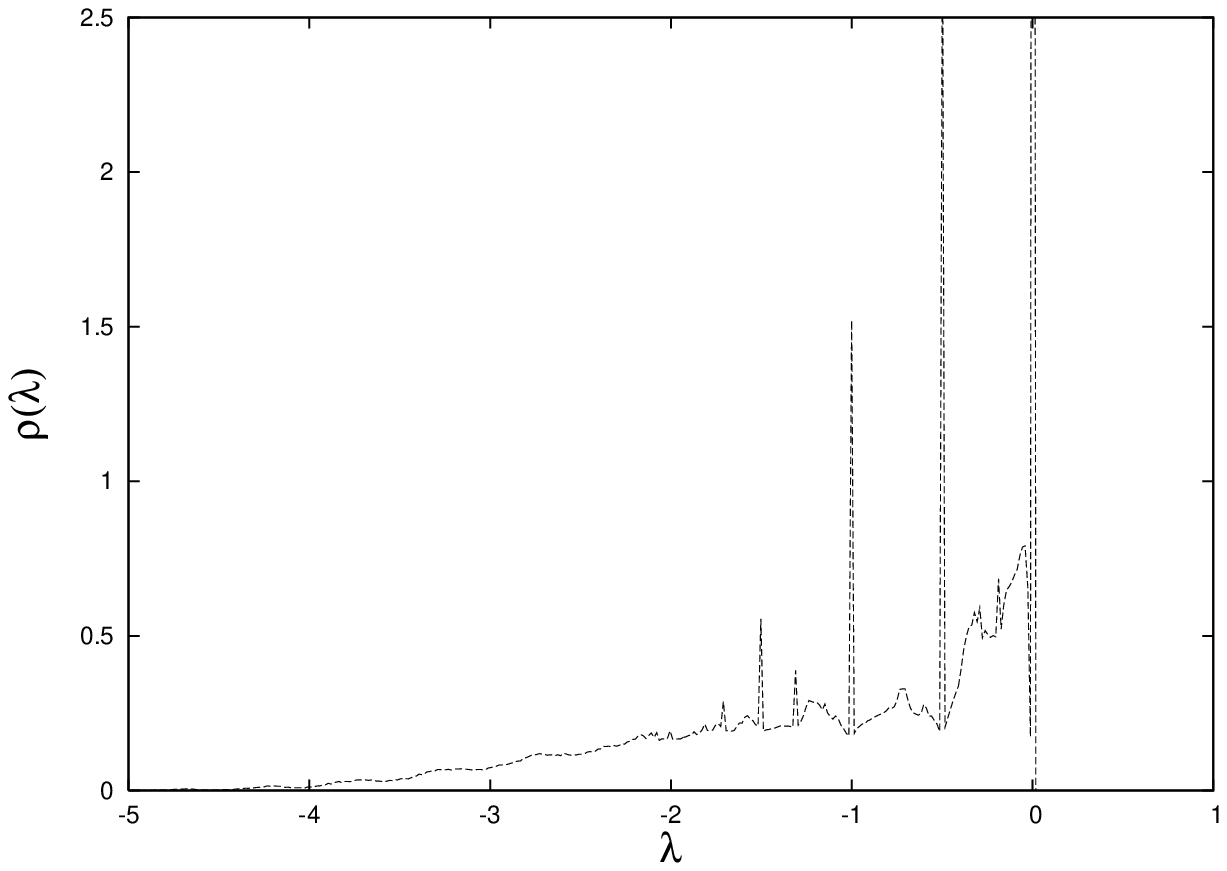,width=0.49\textwidth}
\centering{\hfill\epsfig{file=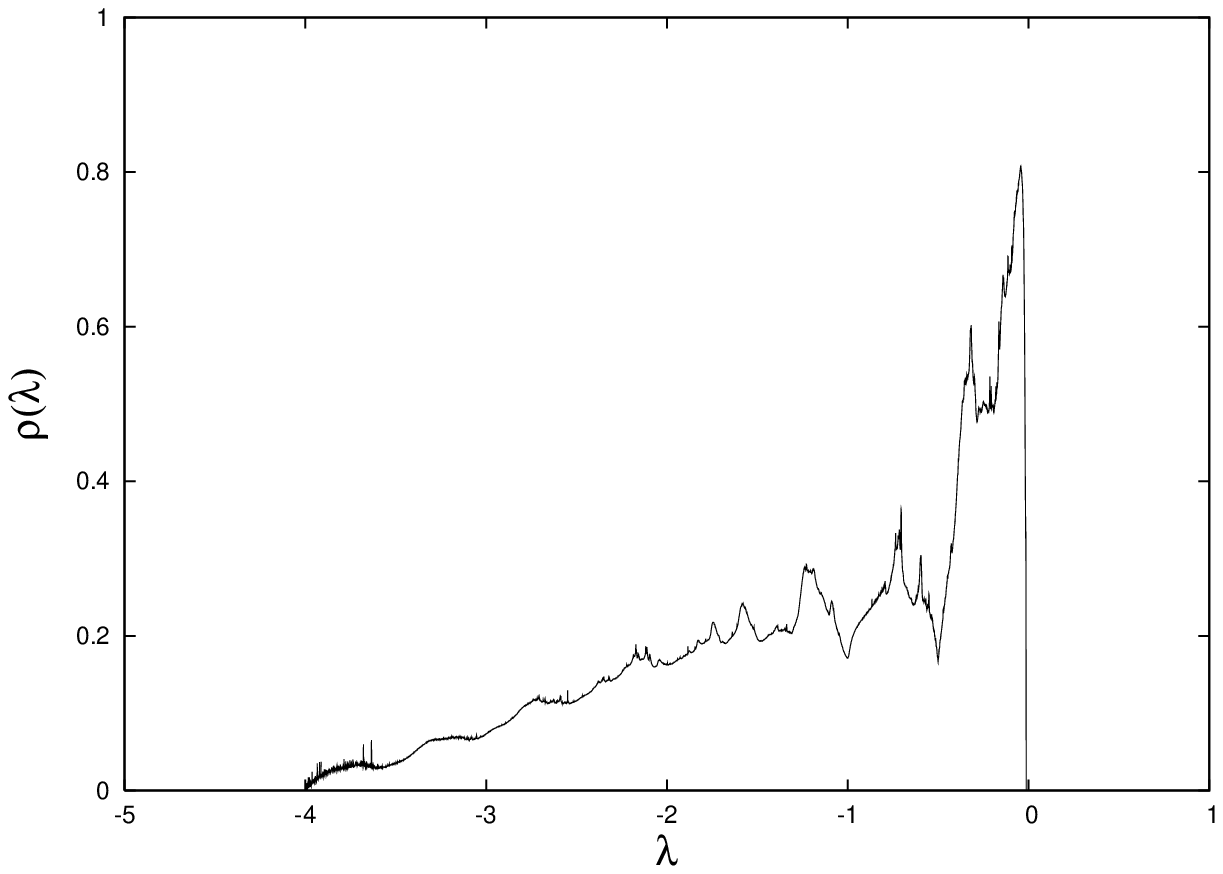,width=0.49\textwidth}}
\caption{Spectral density for the Laplacian on a Poissonian random graph with 
$c=2$ as computed via the present algorithm. Upper left panel: $\varepsilon=
10^{-3}$-results; upper right panel: results from  numerical diagonalisation 
of $N\times N$ matrices of the same type with $N=2000$. Lower panel: continuous
part of the spectrum obtained using $\varepsilon =10^{-300}$ as a regularizer.}
\label{fig:p2L}
\end{figure}

To evaluate the spectral density within the present framework one would thus have
to compute
$$ 
\overline{ Z_N^n} = \int \prod_{ia} \frac{\rd u_{ia}}{\sqrt{2\pi/i}} 
\exp\left\{ -\frac{i}{2}\lambda_\varepsilon \sum_{i,a} u_{ia}^2 
+ \frac{c}{2N} \sum_{ij} \left(\left\langle\exp\Bigg(\frac{i K}{2}\sum_{a} (u_{ia}-u_{ja})^2\Bigg)\right\rangle_K  -1 \right)\right\}
$$
instead of (\ref{avZn}). The required modification has, of course, been noted 
earlier \cite{BrayRodg88,Staer+03}. The resulting problem constitutes precisely 
(the harmonic variant of) the translationally invariant systems, for which the 
framework in \cite{Ku+07} was developed in the first place. The general theory 
can be copied word for word, and the fixed point equations  (\ref{hatpif}), 
(\ref{pif}) remain formally unaltered except for the change in $Z_2(\omega,
\omega',K)$ in (\ref{defZ2s}), owing to the modified interaction term, which 
gives rise to a modified expression for $\hat\Omega(\omega',K)$  in 
(\ref{defhatOm}). We obtain
\be
\hat\Omega(\omega',K)=\frac{K \omega'}{K - i\omega' }
\label{defhatOm2}
\ee
instead of (\ref{defhatOm}). Fig. \ref{fig:p2L} shows the spectrum of a Laplacian
for a Posisson random graph with $c=2$, comparing our solution (upper left panel) 
computed with $\varepsilon=10^{-3}$ with numerical diagonalization results in the 
upper right panel. We use $K_{ij}\equiv 1/c$ for the non-zero matrix elements in 
this case. As in the other cases, we observe a localization transition, here at 
$\lambda_c \simeq -3.98$. Results obtained with a small regularizer $\varepsilon
=10^{-300}$ exhibiting only the continuous part of the spectrum are shown in the 
lower panel.

%----------------------------------------------------------------------------------
\subsection{Unfolding Spectral Densities}
%----------------------------------------------------------------------------------

As a last item in this study we look at the possibility of unfolding the spectral
density according to contributions of local densities of state, coming from vertices
of different coordination, as suggested by Eq. (\ref{spec1}). This method has been 
used in \cite{Ku+07} to look at distributions of Debye-Waller factors in amorphous 
systems, unfolded according to local coordinations. In the present context it may
provide an interesting diagnostic tool to help understanding localization phenomena.

\begin{figure}[ht]
\hfil\epsfig{file=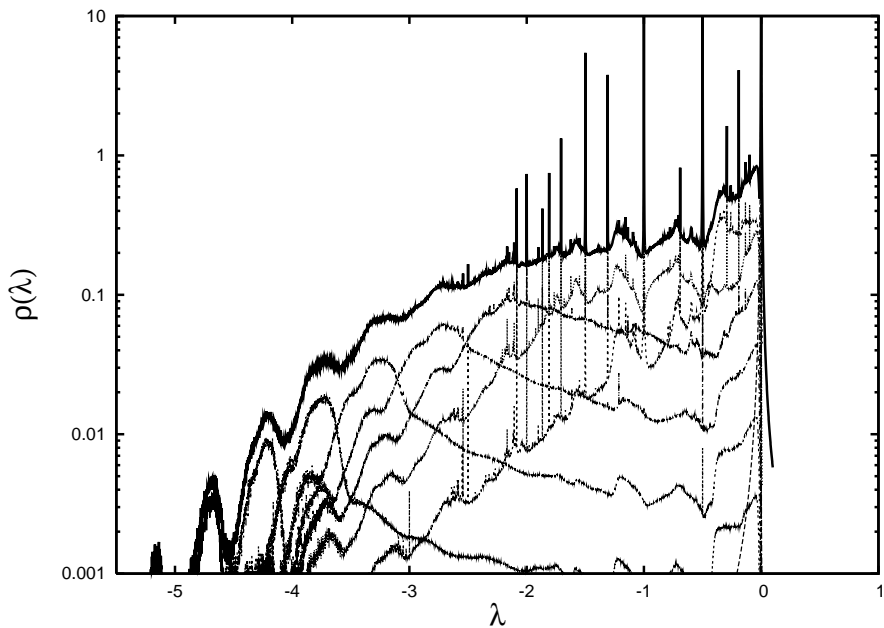,width=0.75\textwidth}\hfill
\caption{Spectral density for the Laplacian on a Poissonian random graph with 
$c=2$ (full upper line), shown together with its unfolding according to contributions 
of different coordination, as discussed in the main text.}
\label{fig:spe-unf}
\end{figure}

Fig \ref{fig:spe-unf} exhibits the spectrum of the graph Laplacian shown in the 
previous figure along with its unfolding into contributions of local densities of 
state with different coordination. The present example clearly shows that --- 
somewhat paradoxically --- the well connected sites are the ones providing the 
dominant contributions to localized states in the lower band-edge Lifshitz tails.
The clearly identifiable humps in the figure correspond from left to right to $k=9$,
$k=8$, $k=7$, $k=6$, $k=5$, $k=4$, and $k=3$, which easily allows to identify the 
corresponding contributions to the spectral density, the contribution of $k=2$ 
gives rise to several notable humps in the spectral density, and together with the $k=1$
contribution is mainly responsible for the dip at $\lambda=-1$. The $k=0$ contribution
is mainly responsible for the $\delta$-peak at $\lambda=0$ (which is broadenend into 
a Lorentzian of width $\varepsilon=10^{-3}$ due to the regularization, as discussed
earlier.

%----------------------------------------------------------------------------------
\section{Conclusions}
%----------------------------------------------------------------------------------

In the present paper we have used a reformulation of the replica approach to the
computation of spectral densities for sparse matrices, which allows to obtain spectral
densities in the thermodynamic limit to any desired detail --- limited only by 
computational resources. Our method is versatile in that it allows to study systems
with arbitrary degree distributions, as long as they give rise to connectivity 
distributions with finite mean. A cavity approach that emphasises results on finite 
instances will appear elsewhere \cite{Rog+08}. As expected (and well known), the 
Wigner semi-circle reemerges in the large $c$ limit as discussed in \cite{Rog+08}. 
Large and small $\lambda$ asymptotics remain to be investigated. Our method allows 
to expose the separate contributions of localized and extended states to the spectral 
density, and thereby to study localization transitions. We shall explore this issue 
in greater detail in a separate publication. Indeed, with results for graph-Laplacians
in hand, the step towards a study of discrete random Schr\"odinger operators and
Anderson localization in such systems is just around the corner \cite{KuMour08}.
A generalization to asymmetric matrices using both the cavity method and a replica 
approach for the ensemble along the lines of \cite{Haake+92} is currently under
investigation in our group \cite{AnandRog08}. Other problems we have started to look 
at are spectra of modular systems \cite{ErgKu08} and small world networks. 

We believe our results to constitute an improvement over previous asymptotic results 
as well as over results obtained by closed form approximations. They may open the way 
to further interesting lines of research. Let us here mention just a few such examples: 
within RMT proper, one might wish to further investigate the degree of universality of 
level correlations in these systems \cite{MirFyo91}; one could refine the random 
matrix analysis of financial cross-correlations \cite{Plerou+99} by taking non-trivial 
degree distributions of economic interactions into account, or one might wish to look 
at finite connectivity variants of random reactance networks \cite{Fyod99}, taking 
e.g. regular connectivity 4 to compare with results of numerical simulations of such 
systems on two-dimensional square lattices.

{\bf Acknowledgements} It is a pleasure to thank G\"uler Erg\"un, Jort van Mourik, Isaac P\'erez-Castillo, Tim Rogers and Koujin Takeda for illuminating discussions. Jort van 
Mourik also kindly supplied instances of scale free-random graphs to allow comparison 
of ensemble results and results from numerical diagonalization in this case.

\appendix

\section{Population Dynamics}
The stochastic algorithm used to solve (\ref{hatpif}), (\ref{pif}) takes the following
form. Populations $\{\omega_i; 1\le i\le N_p\}$  and $\{\hat\omega_i; 1\le i\le N_p\}$
are randomly initialized with ${\rm Re}~ \omega_i >0$ and ${\rm Re}~ \hat\omega_i >0$.

Then the following steps are iterated
\begin{description}
\item[1.] Generate a random $k \sim \frac{k}{c} p_c(k)$.
\item[2.] Randomly select $k-1$ elements from $\{\hat\omega_i; 1\le i\le N_p\}$;
compute
\be 
\Omega = i\lambda_\varepsilon + \sum_{j=1}^{k-1} \hat\omega_{i_j}\ ,
\ee
and replace $\omega_i$ by $\Omega$ for a randomly selected  $i\in \{1,\dots,N_p\}$.
\item[3.] Select $j \in \{1,\dots,N_p\}$ at random, generate a random $K$ according
to distribution of bond strengths; compute
\be
\hat\Omega = \frac{K^2}{\omega_j}\qquad, \qquad \Bigg (\mbox{\rm or}~~~\hat\Omega =\frac{K \omega_j}{K - i\omega_j } ~~~ \mbox{for zero row-sums} \Bigg)\ ,
\ee
and replace $\hat\omega_i$  by $\hat\Omega$ for a randomly selected $i\in \{1,\dots,N_p\}$. 
\item[4.] return to {\bf 1.}
\end{description}

This algorithm is iterated until populations with stable distributions of
$\{\hat\omega_i; 1\le i\le N_p\}$ and $\{\omega_i; 1\le i\le N_p\}$ are attained.

A variant of this algorithm  when implemented on instances of real graphs generates 
the belief-propagation or cavity equations for this problem, as studied in 
\cite{Rog+08}. It can be derived directly in terms iterative evaluations of (\ref{Zn})
on locally tree-like graphs.

%%*******1*********2*********3*********4*********5*********6*********7********%%
\section*{References}
\bibliography{../../MyBib}
%----------------------------------------------------------------------------------
\end{document}